\font\rm=cmr12
\font\smallrm=cmr10
\font\it=cmti12
\font\bf=cmbx12

\baselineskip=18pt
\rm
\overfullrule=0pt

\noindent \hfil {\bf {New Conformal Gauging and the Electromagnetic Theory
of Weyl}}

\bigskip
\bigskip
\noindent \hfil James T. Wheeler

\noindent \hfil {\it {Department of Physics, Utah State University, Logan,
UT 84322}}

\noindent \hfil jwheeler@cc.usu.edu
\bigskip
\bigskip
\noindent \hfil Abstract
\smallskip
A new 8-dim conformal gauging solves the auxiliary field problem and
eliminates unphysical size change from Weyl's electromagnetic theory.  We
derive the Maurer-Cartan structure equations and find the zero curvature
solutions for the conformal connection. By showing that every one-particle
Hamiltonian generates the structure equations we establish a correspondence
between phase space and the 8-dim base space, and between the action and
the integral of the Weyl vector.  Applying the correspondence to generic
flat solutions yields the Lorentz force law, the form and gauge dependence
of the electromagnetic vector potential and minimal coupling. The dynamics
found for these flat solutions applies locally in generic spaces.  We then
provide necessary and sufficient curvature constraints for general curved
8-dimensional geometries to be in 1-1 correspondence with 4-dimensional
Einstein-Maxwell spacetimes, based on a vector space isomorphism between
the extra four dimensions and the Riemannian tangent space.  Despite part
of the Weyl vector serving as the electromagnetic vector potential, the
entire class of geometries has vanishing dilation, thereby providing a
consistent unified geometric theory of gravitation and electromagnetism.
In concluding, we give a concise discussion of observability of the extra
dimensions.

\vfil
\break

\noindent \hfil {\bf {New Conformal Gauging and the Electromagnetic Theory
of Weyl}}
\smallskip
\bigskip
\noindent \hfil James T. Wheeler

\noindent \hfil {\it {Department of Physics, Utah State University, Logan,
UT 84322}}

\noindent \hfil jwheeler@cc.usu.edu

\bigskip
\noindent 1.  Introduction
\smallskip
In 1918, following immediately on the heels of Einstein's introduction of
general relativity,
Weyl proposed a generalization of Riemanian spacetime structure [1] to
allow for local
changes in the standard of length.  When the gauge field required to insure
this symmetry
has nonvanishing curl, the lengths of vectors change as they undergo
parallel transport.
This nicely completes the picture of Riemann in which vectors are rotated
but not changed
in length.  In Weyl geometry, all properties of transported vectors -
magnitude {\it
{and}} direction - are subject to the transport law.

Weyl identified the gauge field of local dilations with the electromagnetic
potential, thereby
explaining the gauge freedom of that potential and giving a geometric
interpretation of
electromagnetic forces.  Unfortunately, the theory is inconsistent with
experience because it
predicts changes in the sizes of objects depending on their paths.
Applied, for example, to
the case of atoms moving in a background electromagnetic field, the
predicted size changes
would produce substantial unobserved broadening of atomic spectral lines.

The simplicity and naturalness of Weyl geometry together with the promise
of a geometric
understanding of electromagnetism makes the failure of Weyl's physical
theory puzzling.
Indeed, many physicists [4-22] including Weyl [1-3] proposed alternative
versions of
the physical interpretation in attempts to make the theory consistent.
Ultimately, the work
led to the current U(1) gauge theory of electromagnetism, which maintains
much of the
beauty of the original proposal without inconsistent predictions such as
changes in atomic
spectra.

In the present work, we show that a new conformal gauging avoids the
standard objection
[5] to this eighty-year-old theory of Weyl.  Since the conformal group
contains the Weyl
group\footnote{$^{1}$}{{\smallrm {Weyl's original theory was based on the
homogeneous Weyl group, consisting of Lorentz transformations plus
dilations.  We will
also require the inhomogeneous Weyl group, which in addition to the
homogeneous
transformations includes four translations.}}}, the spaces resulting from
its gauging share
certain properties with Weyl's original gauge theory of electromagnetism.
However, the 8-dimensional character of the new gauging and its
interpretation as a generalization of phase
space eliminate the problem of unphysical size change.  Before discussing
these new
conformal spaces, which we call {\it {biconformal geometries}}, we review
some results
from previous studies of the conformal group.

The conformal group is the most general set of
transformations preserving ratios of infinitesimal
lengths.  On 4-dim spacetime this group is 15 dimensional.
 Besides Lorentz transformations, it includes translations,
 inverse translations\footnote{$^{2}$}{{\smallrm {The inverse translations
have
alternatively been called special conformal transformations, conformal
boosts, accelerations
or elations.  We choose the term inverse translations because these
transformations are pure
translations at infinity. This is easily seen by translating the inverse
coordinate $y^{\mu} =
- {x^{\mu} \over x^{2}} \rightarrow y^{\mu} + a^{\mu}$} then inverting
again.}} and
dilations.  Considerable attention
[23-27] has been given to 4-dim spacetimes which result
 from gauging this group.  In addition to the usual metric and
 local Lorentz structures of spacetime, these models are expected,
 {\it {a priori}}, to possess five gauge vectors - one for dilations and
 four for the inverse translations.  However, it was long thought
that the inverse translations could not be gauged, because the
 corresponding current is position dependent.  From a slightly
different perspective, Mansouri et {\frenchspacing {al. [26, 27]}}
argue that making the inverse translations local is redundant
with general coordinate transformations so that the corresponding
 gauge fields have no new effect.  Nonetheless, Crispim Rom\~{a}o,
Ferber and Freund [23, 24], and independently Kaku, Townsend and
Van Nieuwenhuizen [25]  show how to gauge the inverse
translations, but find them algebraically removable
from the problem.  This problem is in fact generic:
in any 4-dimensional
scale-invariant torsion-free field theory with action built quadratically
from the curvatures
of the conformal group these four
inverse translational guage fields may be algebraically removed
from the problem [28].  The local inhomogeneous Weyl symmetry of such
theories always
has an effective field theory equivalent to one based on the homogeneous
Weyl group,
and therefore sheds no light on Weyl's theory.

By contrast, consideration of the of the fixed points of the various conformal
transformations suggests the use of an 8-dim base space instead of a 4-dim
base space,
corresponding to the existence of eight distinct translational generators
in the Lie algebra.
Requiring an 8-dim base space makes the gauge fields of the inverse
translations act as part
of the solder form on the base space, so these fields {\it {necessarily}}
remain
independent.  Even in the case of a fully flat biconformal space this
solder form contains
nontrivial physical information.

Our present investigation of this new conformal gauging focuses principally
on the study of
flat biconformal spaces, because this is where the connection with
well-known systems should be most evident and because the dynamics of flat
spaces will also apply locally in generic spaces.  Unlike Riemannian
geometry, there is an entire nontrivial class of
flat solutions. We first derive this class of solutions.  Next, we show
that the flat spaces
possess a symplectic structure which allows their interpretation as
one-particle phase spaces
and provides a concise description of the corresponding Hamiltonian
dynamics.  Applying
this Hamiltonian correspondence to the most general flat solution gives a
geometric
derivation of the Lorentz force law, in strong contrast to the original
Weyl theory.  After
indicating briefly how the biconformal electromagnetic theory connects
satisfactorily to
U(1) gauge theory, we show how to introduce Einstein gravitation, arbitrary
additional
fields and electromagnetic sources into the biconformal structure.

The layout of the paper is as follows.  In Sec.(2) we develop
biconformal structure as a conformal fiber bundle.  We give
particular attention to the general proceedure for building connections
on manifolds based on an underlying symmetry group [29-32],
because it is this proceedure which motivates our use of an 8-dimensional
 base manifold.  The construction consists of two steps.  First, an elementary
geometry is found by taking the quotient of a given symmetry group by a
non-normal subgroup.  Then, the connection of the resulting principal
fiber bundle is generalized by including curvature.  As an example we
 review the construction of general spacetimes by gauging the Poincar\'{e}
 group [33-36].  We then  implement these techniques to gauge the conformal
 group in a new way.  First, we construct an elementary geometry as the
quotient, ${\cal {C}}/ {\cal {C}}_{0}$, of the conformal group, $\cal {C}$,
 by its isotropy subgroup, ${\cal {C}}_{0}$.  This gives a conformal
Cartan connection on an 8-dimensional manifold.  Next, we generalize
the connection to that of a curved 8-dimensional manifold with the
 7-dimensional homogeneous Weyl group as fiber by the addition of horizontal
curvature 2-forms to the group structure equations.  The resulting
8-dimensional
 base manifold is called a {\it {biconformal space}} and the full fiber
bundle the
biconformal bundle.

Our first principal result is the flat solution presented in Sec.(3). We
find a frame field satisfying the zero-curvature Maurer-Cartan structure
equations of biconformal space.

In Sec.(4) and Sec.(5) we establish our next claim:  the classical
Hamiltonian dynamics of a
single point particle is equivalent to the specification of a 7-dim surface
in flat biconformal
space.  As a consequence of the form of the structure equations, there
necessarily exists a
set of preferred curves satisfying the Hamiltonian equations of motion.
This claim is
established in two steps.  First, in Sec.(4), we
show that the classical Hamiltonian description of a point particle
defines a class of 8-dim differential geometries with structure equations
of manifestly biconformal type.  Then in Sec.(5) we show that there is an
embedding,
unique up to local symplectic changes of basis, of the Hamiltonian system
into a {\it {flat}}
biconformal geometry.  We also demonstrate the necessary existence of a
preferred set of
curves in the biconformal space satisfying the Hamiltonian equations
of motion.  The unique local equivalence between hypersurfaces in flat
biconformal
geometry and Hamiltonian systems provides a clear physical interpretation
of the geometric variables of biconformal space.

The central importance of this second result is in
definitively establishing the physical interpretation
of the new conformal gauging.  Our main goal is not to provide an alternate
formulation of
classical mechanics, but rather to use this embedding as a guide to subsequent
physical interpretation of the elementary biconformal variables.  Given the
result of
Sec.(4), that a classical single particle Hamiltonian system generates a
class of biconformal
spacetimes and its converse in Sec.(5), we can conclude that for an
isolated test particle in a general biconformal space the four new
coordinates may be interpreted locally as the corresponding generalized
particle momentum. Our identification therefore provides a physical
correspondence principle for biconformal spaces.  While in general spaces
the extra dimensions will not necessarily represent momentum globally, we
can always take the limit of a tightly confined field in a local Lorentz
frame, for which the extra dimensions will permit the momentum
interpretation.  Indeed, when non-flat biconformal spaces are investigated
in Sec.(7), we will see that the ``momentum-like" co-solder form contains
the stress-energy tensor of any gravitational sources.  This generalization
from momentum to the stress-energy tensor is exactly what one would
anticipate in moving from a particle interpretation to a field
interpretation.

For our third result, in Sec.(6), we apply the Hamiltonian correspondence
of the previous
sections to the general solution for flat biconformal space, to show how
that flat solution
predicts the following properties of the electromagnetic vector potential:
\smallskip
\item{1.} The 4-vector form of the potential.  This is nontrivial, since it
involves the
reduction of the 8-dim 1-form $\omega_{0}^{0} = \omega_{0\mu}^{0}(x, y){\bf
{d}}x^{\mu} + \omega_{0}^{0\mu}(x, y){\bf {d}}y_{\mu}$ to a 4-dim 1-form,
$\alpha_{\mu} (x) {\bf {d}}x^{\mu}$ on spacetime.
\item{2.} Its usual gauge dependence $\alpha'(x) = \alpha(x) + \bf {d}\phi$.
\item{3.} Minimal coupling $p_{a} \longrightarrow p_{a} - \lambda \alpha_{a}$.
\item{4.}The correct equation of motion for a charged particle moving under
its influence.
\smallskip
{\it {This section therefore provides a consistent realization of Weyl's
goal [1] of
expressing electrodynamics in terms of the dilational gauge field}}.  Upon
moving to more general biconformal spaces, the remarks of the preceeding
paragraph imply that in a local Lorentz frame an isolated, charged test
particle will move according to the Lorentz force law.

These predictions of electromagnetic effects in biconformal space follow
directly from the
relationship of biconformal space to phase space and Hamiltonian dynamics,
developed in
Secs.(4) and (5).  No further assumption is necessary to derive the Lorentz
force law.  At
the same time, the biconformal theory predicts constancy of size.  There is
an entire class of
{\it{flat}} biconformal spaces which includes sufficient freedom in the
guage fields to
account for the classical electrodynamics of a charged point particle in an
arbitrarily specified background electromagnetic field.  The flatness
condition includes vanishing dilational curvature, so none of these
geometries leads to size change of any kind.  In sharp contrast to Weyl
geometry in which vanishing dilation implies vanishing gauge vector, the
dilational gauge vector in flat biconformal space is {\it {required to be
nonzero}}.

Sec.(6) ends with a brief comparison of the biconformal model with Weyl's
original gauge
theory and with the standard U(1) model of electromagnetism.

In Sec.(7), we show how to introduce arbitrary gravitational and
electromagnetic
sources so that the results of Sec.(6) remain valid in the resulting class
of curved
biconformal spaces.  The entire class has vanishing dilation, and the
electromagnetic and
gravitational fields satisfy the Maxwell and Einstein equations,
respectively. {\it {This section therefore provides a consistent
realization of Weyl's goal of a unified geometric theory of gravitation and
electromagnetism.}}  While other fundamental interactions have been
discovered since Weyl's time, the inclusion of both of the large scale
forces in a geometric theory must be considered a step in the right
direction; moreover, biconformal spaces contain additional fields which are
not studied here.  These additional fields might have an interpretation as
the additional interactions.  In any case, there is always the possiblilty
of adding internal symmetries beyond the conformal symmetry.

The final section consists of a detailed discussion of possible
consequences of regarding energy and momentum variables as four of the
coordinates in an 8-dimensional space.  We consider the necessary
isomorphism of mathematical structure, transformation properties and
dynamical laws that must hold between momentum space variables and the
biconformal co-space variables.  Collisions and interactions are discussed
with attention given to continuity and the proper proximity of colliding
particles.  Finally, we point to some experimental results which suggest a
coordinate-like behavior of momentum variables.
\bigskip
\noindent 2.  The construction of spacetimes with local symmetries
\smallskip
Before defining and interpreting biconformal spaces, we require some
background
motivating their construction.  While it is customary to begin the
discussion of a differential
geometry with the specification of a manifold and metric pair, $({\cal
{M}}, g)$, our
present interest lies in developing a spacetime model beginning with
locally determined
symmetry considerations.  This allows us to construct geometries which {\it
{a priori}}
possess specified local symmetries.  It is in part the elements of this
construction which
motivate our choice of an 8-dimensional base manifold for conformal gauging
instead of
the usual 4-dimensional picture.  This in turn requires us to interpret the
extra four
dimensions.  The interpretation of the extra coordinates in the zero
curvature limit is
accomplished in Secs.(4-6).

We consider symmetry groups which include
the local Lorentz symmetry of spacetime, seeking a
procedure for developing a connection on a spacetime
$({\cal {M}}, g)$ from a knowledge of an experimentally
determined (i.e.``global") symmetry group.  We use the well-known
techniques of Cartan and Klein [29-32, 37-39], proceeding as follows.
We begin with a Lie group, ${\cal {G}}$, and a subgroup
${\cal {G}}_{0}$ called the isotropy subgroup.  The isotropy
subgroup should contain no normal subgroup of ${\cal {G}}$ if the full group
is to act effectively and transitively on the base space.  Defining
a projection onto a base space as the quotient
${\cal {G}}/{\cal {G}}_{0}$ the group manifold becomes
a fiber bundle, called by Klein an {\it {elementary geometry}}.
The full group will act effectively and transitively on the base
space of the elementary geometry.  The base space will be a manifold if and
only if the
fibration is regular, i.e., there exists a neighborhood of each point of
any fiber which that
fiber intersects only once.  This regularity condition holds for the groups
we consider. This
base space of the
elementary geometry therefore provides a manifold upon which we
now generalize the connection to a Cartan connection by introducing curvature.

The generalization to a Cartan connection occurs by only requiring the action
 of the full group to be detectable on curves instead of globally.
Specifically,
we may introduce into the structure equations any curvature
2-forms consistent with the resulting Bianchi identities and the
following requirement.  Let ${\cal {P}}(\lambda)$ be any curve in the
bundle, let
$\omega_{j}^{i}$ be a connection on the bundle,
and let $f: {\cal {G}} \longrightarrow V^{N^{2}}$ be a linear representation
of ${\cal {G}}$ by $N \times N$ matrices.  Then integrating
$$ {\bf {d}}f_{j}^{i} = \omega_{j}^{k} f_{k}^{i}$$
along ${\cal {P}}(\lambda)$ yields a group transformation
$f_{j}^{i}(\lambda_{0})$ at
each
point ${\cal {P}}(\lambda_{0})$.  The connection $\omega_{j}^{i}$ will be a
Cartan
connection if the transformation $f_{j}^{i}(\lambda)$ depends only on the
projection of
the curve ${\cal {P}}(\lambda)$ into the base space.  This condition holds
if and only if
the curvatures are {\it {horizontal}}, i.e., bilinear in those
connection forms which vanish on the fibers.  The generalization
to a curved connection occurs because different curves
in the base space between the same pair of points can give
different group elements.

If the connection is linear we can express this condition in terms of
mappings of
orthonomal frames.  Consider again a curve ${\cal {P}}(\lambda)$, an
initial point
${\cal {P}}(0)$ on the curve, and a frame ${\cal {E}}_{a}(0)$ at the point.
We then can demand that there shall exist a transformation from the full
group ${\cal {G}}$ mapping the initial pair $({\cal {P}}(0), {\cal
{E}}_{a}(0))$ to a
corresponding pair $({\cal {P}}(\lambda), {\cal {E}}_{a}(\lambda))$ for any
point ${\cal {P}}(\lambda)$ on the curve.

We illustrate the method using the Poincar\'{e} group.  The group may be
described locally
by its structure equations (equivalent to its Lie algebra), which take the form
$$\eqalignno{
{\bf {d}}\omega_{b}^{a} &=  \omega_{b}^{c} \wedge \omega_{c}^{a}  \cr
{\bf {d}}\omega^{a} &=  \omega^{b} \wedge \omega_{b}^{a}  &(2.1) \cr}$$
where the one-forms $\omega_{b}^{a}, \omega^{a} \enskip (a, b = 1, 2, 3,
4)$ span the
10-dim group manifold.
The only subgroup ${\cal {G}}_{0}$ containing no normal subgroup of ${\cal
{G}}$ but which does include the Lorentz group is the Lorentz group itself,
since the translations form a normal subgroup.  The quotient ${\cal
{G}}/{\cal {G}}_{0}$ leads to a 4-dim base space with Lorentz fibers.  It
is easy to see that the base space is Minkowski space, and therefore the
elementary geometry is the bundle of orthonormal frames over Minkowski
spacetime.

Continuing with the Lorentz bundle, we alter the connection to that of a
curved base space
by adding curvature 2-forms, $\Omega_{b}^{a}$ and $\Omega^{a}$, to the
structure
equations.  Clearly  there is one curvature 2-form for each generator of
the original Lie
group.  Furthermore, each curvature component is horizontal, depending only
on the forms
$\omega^{a}$ which vanish on the fibers.  Functionally, the curvature
components depend only on the quotient space, ${\cal {G}}/{\cal {G}}_{0}$.
Thus,
$$\eqalignno{
{\bf {d}}\omega_{b}^{a} &=  \omega_{b}^{c} \wedge \omega_{c}^{a} +
\Omega_{b}^{a} \cr
{\bf {d}}\omega^{a} &=  \omega^{b} \wedge \omega_{b}^{a} + \Omega^{a} &(2.2)
\cr}$$
or, in more familiar notation,
$$\eqalignno{
{\bf {R}}_{\enskip b}^{a} &= {\bf {d}} \omega_{b}^{a}- \omega_{b}^{c} \wedge
\omega_{c}^{a} \cr
{\bf {T}}^{a} &= {\bf {de}}^{a}- {\bf {e}}^{b} \wedge  \omega_{b}^{a}  &(2.2')
\cr}$$
where $\omega_{b}^{a}$ is the spin connection, ${\bf {e}}^{a}$ the
vierbein, ${\bf
{R}}_{\enskip b}^{a}$ the curvature 2-form and ${\bf {T}}^{a}$ the torsion.
The horizontality condition guarantees that the generator of a Poincar\'{e}
transformation
found by integrating the connection 1-forms along any curve ${\cal
{P}}(\lambda)$ will
depend only on the projection of the curve into the base manifold, $\pi({\cal
{P}}(\lambda))$.  Specifically, define
$$\eqalignno{
{\bf {d}}_{1}{\cal {P}} &= \omega^{a} {\cal {E}}_{a}  \cr
{\bf {d}}_{1}{\cal {E}}_{a} &= \omega_{a}^{b} {\cal {E}}_{b} &(2.3) \cr}$$
as the change in the point ${\cal {P}}$ and the vector frame ${\cal
{E}}_{b}$ along any
curve in the bundle, where ${\bf {d}}_{1}$ is the 1-dim exterior derivative
on the curve.
It is convenient to introduce a full 10-dim frame $({\cal {E}}_{a}, {\cal
{F}}_{a}^{b})$
at the initial point, with the frame ${\cal {E}}_{a}$ being the horizontal
part.  We choose
$({\cal {E}}_{a}, {\cal {F}}_{a}^{b})$ to be dual to $(\omega^{a},
\omega_{b}^{a}):$
$$\eqalignno{
\omega^{a}({\cal {E}}_{b}) &= \delta_{b}^{a} \quad \omega^{a}({\cal
{F}}^{b}_{c})
= 0  \cr
\omega^{a}_{b}({\cal {E}}_{c}) &= 0  \quad \omega^{a}_{b}({\cal
{F}}_{d}^{c}) =
\delta_{d}^{a}\delta_{b}^{c} &(2.4) \cr }$$
When eqs.(2.3) are integrated along a curve ${\cal {P}}(\lambda)$ we find a
new point-frame pair at each value of $\lambda$.  Now let ${\cal
{P}}(\lambda)$ be the closed
perimeter of an arbitrary infinitesimal plaquette with area element S.
Then explicit
evaluation of the integral of ${\bf {d}}_{1}{\cal {P}}$ around the
plaquette gives:
$$ \oint_{\cal {P}} {\bf {d}}_{1}{\cal {P}} \approx   \Omega^{a}(S){\cal
{E}}_{a}(0)
\eqno (2.5)  $$
A general infinitesimal surface element S may be expanded in terms of the
full frame at the
initial point as
$$ S = S^{ab}{\cal {E}}_{a} \wedge {\cal {E}}_{b} + S^{ab}_{c}{\cal
{F}}_{a}^{c}
\wedge {\cal {E}}_{b} + S^{ab}_{cd}{\cal {F}}_{a}^{c} \wedge {\cal
{F}}_{b}^{d}.
\eqno(2.6)$$
Now taking $\Omega^{a}$ horizontal, $\Omega^{a} = \Omega^{a}_{bc} \enskip
\omega^{b} \wedge \omega^{c}$ we use eqs.(2.4) to evaluate
$$\eqalignno{
\Omega^{a}(S) &= \Omega^{a}_{bc} [S^{de} \omega^{b}({\cal {E}}_{d})
\omega^{c}({\cal {E}}_{e}) +S^{de}_{f} \omega^{b}({\cal {F}}_{d}^{f})
\omega^{c}({\cal {E}}_{e}) +S^{de}_{fg} \omega^{b}({\cal {F}}_{d}^{f})
\omega^{c}({\cal {F}}_{e}^{g})] \cr
&= \Omega^{a}_{bc} S^{de} \delta^{b}_{d} \delta^{c}_{e} \cr
&= \Omega^{a}(\pi(S)) &(2.7) \cr}$$
so that the path dependence of the point ${\cal {P}}(\lambda)$ depends only
on the
projection of the loop into the base manifold.  A completely analogous
argument holds for
the integral of ${\cal {E}}_{a}$ and the horizontality of $\Omega_{b}^{a}$.

Eqs.(2.2) or ($2.2'$) now describe a curved 4-dim spacetime
 with local Lorentz symmetry.  The second curvature,
  $\Omega^{a} = {\bf {T}}^{a}$ allows the inclusion of
 torsion.  Because the remaining Lorentz symmetry of the
fibers does not mix the components of the torsion with the
components of the Riemann curvature, the usual specification
of general relativity, $\Omega^{a} = 0$, is consistent.  General
 relativity also requires identification of the cotangent space
$T^{*}$ with the space spanned by the solder form ${\bf {e}}^{a}$.
This identification is automatic here because ${\bf {e}}^{a}$ is required
${\it {a priori}}$ to span the base space ${\cal {G}}/{\cal {G}}_{0}$.  As
we have
carried out the construction here, the topology of the manifold is that of
the quotient, $R^{4}$, but nontrivial base spaces are easily achieved by
specifying an arbitrary manifold ${\cal {M}}$ having the same (Minkowski)
cotangent
space.  This substitution of ${\cal {M}}$ for ${\cal {G}}/{\cal {G}}_{0}$
is allowed because it alters only topological properties, while all of the
structures of interest described above are local.

Understanding a symmetry-based approach has significant advantages as we
now turn to
our examination of the conformal group.  The essential point is that, in
contrast to the
Poincar\'{e} example, the presence of inverse translations means that there
are no normal
subgroups, making the choice of isotropy subgroup ${\cal {C}}_{0}$
nontrivial.  While
biconformal space is based on the choice of the homogeneous Weyl group as
${\cal
{C}}_{0}$, it is instructive to first consider what happens if we use the
inhomogenous
Weyl group instead.

As described in the introduction, considerable attention [23-27] has been
given to models
which result from the choice of the 11-dim inhomogeneous Weyl group as ${\cal
{C}}_{0}$.  This choice is natural enough since it gives the same base
manifold as in the
Poincar\'{e} case.
It would appear that this model simply extends the Lorentz fiber symmetry
 of the local Poincar\'{e} model to include dilations and inverse
 translations.  However, in generic field theories based on this bundle
the four components of the
connection corresponding to the inverse translations are generically
auxiliary and may be algebraically removed
from the problem [28].  The resulting field theory, equivalent to
one based on the homogeneous Weyl group, has had at best mixed
success as a field theory.  In any case, no new symmetries or
additional physical fields have been gained in passing from the Weyl
group to the full conformal group.  By contrast, choosing ${\cal {C}}_{0}$
to be the {\it
{homogeneous}} Weyl group from the start, we retain the full 15 degrees of
freedom
 of the original group and acquire many new fields, with the simplest
 example of a biconformal space having a natural physical interpretation
 in terms of Hamiltonian dynamics.

We arrive at the biconformal model if we ask what property of the Lorentz
transformations
made them a suitable fiber symmetry the Poincar\'{e} group.  The
requirement that the
isotropy subgroup should have no normal subgroup ruled out the
translations, which form
a normal subgroup of the Poincar\'{e} group.  The Lorentz symmetry is
therefore the only
possible fiber symmetry.
But there is another way to see that we should use the Lorentz group as the
isotropy
subgroup, based on fixed points.  On Minkowski space, the class of
translations has no
fixed points, while the class of Lorentz transformations leaves the origin
fixed (hence the
name, ``isotropy subgroup").  Thus, we can distinguish the isotropy
subgroup of the
fibering from the ``translational" symmetry of the base manifold by
counting fixed points.

Returning to the conformal group, we find that when we distinguish the
conformal
transformations based on their fixed points, there are simply eight
translations
acting on {\it {compactified}} Minkowski space [40].  A special point, its
null cone, and
an ideal 2-sphere are added at infinity to accomplish the compactificaton.
As a result, the
translations are no longer characterized by an absence of fixed points.
Instead, the class of
translations and the class of inverse translations each has a single fixed
point (the origin and
the point at infinity, respectively) while the Lorentz subgroup and the
dilational subgroup
leave both of these points fixed.  The dilations also leave the ideal
2-sphere fixed.

With these observations, we take the isotropy subgroup ${\cal {C}}_{0}$ to
be the 7-dim
homogeneous Weyl group, consisting of the six Lorentz transformations
together with
dilations.  Defining a projection as the quotient ${\cal {C}}/{\cal
{C}}_{0}$ we are led to
a fiber bundle with an 8-dim manifold as the base space and the homogeneous
Weyl group
as a typical fiber.  This will break {\it {both}} the translational and
inverse translational
symmetries when the base manifold becomes curved, but we nonetheless retain
the full 15
guage field degrees of freedom.

We choose the $O(4,2)$ representation of the conformal
group for our notation [40], with $(A, B, \dots) = (0, 1, \ldots, 5)$.
Letting boldface or
Greek symbols denote forms and $(a, b, \ldots) = (1, \ldots, 4)$, the
$O(4,2)$ metric is
given by $\eta_{ab} = diag(1, 1, 1, -1)$ and $\eta_{05} = \eta_{50} = 1$
with all other
components vanishing.  Introducing the connection 1-form $\omega_{B}^{A}$,
we may
express the covariant constancy of $\eta_{AB}$, as
$${\bf {D}}\eta_{AB} \equiv {\bf {d}}\eta_{AB} - \eta_{CB} \omega_{A}^{C} -
\eta_{AC} \omega_{B}^{C} = 0. \eqno (2.8) $$
We may break the connection form into four independent Weyl-invariant
parts:  the {\it
{spin connection}}, $\omega_{b}^{a}$, the {\it {solder form}},
$\omega_{0}^{a}$, the
{\it {co-solder form}}, $\omega_{a}^{0}$, and the {\it {Weyl vector}},
$\omega_{0}^{0}$ where the spin connection satisfies
$$\omega_{b}^{a} = - \eta_{bc} \eta^{ad} \omega_{d}^{c}  \eqno (2.9a) $$
and the remaining components of $\omega_{B}^{A}$ are given in terms of these by
$$ \eqalignno{
\omega_{0}^{5} &=  \omega_{5}^{0} = 0 &(2.9b) \cr
\omega_{5}^{5} &= - \omega^{0}_{0}   &(2.9c)  \cr
\omega_{5}^{a} &= - \eta^{ab} \omega_{b}^{0} &(2.9d)  \cr
\omega_{a}^{5} &= - \eta_{ab} \omega_{0}^{b}  & (2.9e) \cr } $$

\noindent   These constraints reduce the number of independent fields
$\omega_{B}^{A}$
to the required 15 and allow us to restrict $(A, B, \dots) = (0, 1, \ldots,
4)$ in all
subsequent equations.  The structure constants of the conformal Lie algebra
now lead
immediately to the Maurer-Cartan structure equations of the conformal
group.  These are
simply
$${\bf {d\omega}}_{B}^{A} = {\bf {\omega}}_{B}^{C} \wedge {\bf
{\omega}}_{C}^{A} \eqno (2.10) $$
When broken into parts based on homogeneous Weyl transformation properties,
eq.(2.10)
gives:
$$\eqalignno{
{\bf {d\omega}}_{b}^{a} &= \omega_{b}^{c} \wedge \omega_{c}^{a} + {\bf
{\omega}}_{b}^{0} \wedge {\bf {\omega}}_{0}^{a} - \eta_{bc}\eta^{ad} {\bf
{\omega}}_{d}^{0} \wedge {\bf {\omega}}_{0}^{c} \cr
{\bf {d\omega}}_{0}^{a} &= {\bf {\omega}}_{0}^{0} \wedge {\bf
{\omega}}_{0}^{a}
+ {\bf {\omega}}_{0}^{b} \wedge {\bf {\omega}}_{b}^{a} \cr
{\bf {d\omega}}_{a}^{0} &= {\bf {\omega}}_{a}^{0} \wedge {\bf
{\omega}}_{0}^{0}
+  {\bf {\omega}}_{a}^{b} \wedge {\bf {\omega}}_{b}^{0} \cr
{\bf {d\omega}}_{0}^{0} &= {\bf {\omega}}_{0}^{a} \wedge {\bf
{\omega}}_{a}^{0}
&  (2.11) \cr} $$
Since no finite translation can reach the point at infinity and no inverse
translation can reach
the origin, the space ${\cal {C}}/{\cal {C}}_{0}$ gives a copy of (noncompact)
Minkowski space for each of the two sets of translations.  Since the
generators of the two
types of translation commute modulo the homogeneous Weyl group, the full
base manifold
is simply the Cartesian product of these two Minkowski spaces.  The
generalization to a
curved base space is immediate.  We have:
$$\eqalignno{
{\bf {d\omega}}_{b}^{a} &= \omega_{b}^{c} \wedge \omega_{c}^{a} +
\omega_{b}^{0} \wedge \omega_{0}^{a} - \eta_{bc}\eta^{ad} \omega_{d}^{0}
\wedge
\omega_{0}^{c} + \Omega_{b}^{a} &(2.12a) \cr
{\bf {d\omega}}_{0}^{a} &= \omega_{0}^{0} \wedge \omega_{0}^{a} +
\omega_{0}^{b} \wedge \omega_{b}^{a} + \Omega_{0}^{a} &(2.12b) \cr
{\bf {d\omega}}_{a}^{0} &= \omega_{a}^{0} \wedge \omega_{0}^{0} +
\omega_{a}^{b} \wedge \omega_{b}^{0} + \Omega_{a}^{0} &(2.12c) \cr
{\bf {d\omega}}_{0}^{0} &= \omega_{0}^{a} \wedge \omega_{a}^{0} +
\Omega_{0}^{0}  &  (2.12d) \cr} $$
We will call the four types of curvature $\Omega_{b}^{a}, \Omega_{0}^{a},
\Omega_{a}^{0}$ and $\Omega_{0}^{0}$ the Riemann curvature, torsion,
co-torsion and
dilational curvature, respectively.  Notice that if we set $\omega_{a}^{0}$,
$\omega_{0}^{0}$ and the corresponding curvatures to zero, we recover
eqs.(2.2) for a
4-dim spacetime with Riemannian curvature $\Omega_{b}^{a}$ and torsion
$\Omega_{0}^{a}$.  If we set only $\omega_{a}^{0} =  \Omega_{a}^{0} = 0$, the
structure equations are those of 4-dim Weyl geometry.

Horizontality requires each of the curvatures to take the form
$$\Omega_{B}^{A} = {\scriptstyle {1 \over 2}} \> \Omega_{Bcd}^{A} \>
\omega_{0}^{c} \wedge \omega_{0}^{d} + \Omega_{Bd}^{Ac} \> \omega_{0}^{d}
\wedge \omega_{c}^{0} + {\scriptstyle {1 \over 2}} \> \Omega_{B}^{Acd}\>
\omega_{c}^{0} \wedge \omega_{d}^{0} \eqno (2.13) $$
Based on the interpretation of biconformal space as a generalization of
phase space
(Secs.(4) and (5) below, see also [41]) we will call $\Omega^{A}_{Bcd}$ the
spacetime
term, $\Omega^{Ac}_{Bd}$ the cross term and $\Omega^{Acd}_{B}$ the momentum
term of each type of curvature.  In sharp contrast to the 4-dim gauging, in
which the fiber
symmetry mixes the curvatures $\Omega_{b}^{a}$, $\Omega_{0}^{a}$,
$\Omega_{a}^{0}$, and $\Omega_{0}^{0}$ (see [42]), our choice of the
homogeneous Weyl group
as the structure group not only leaves these independent, but also does not
mix the
spacetime, cross or momentum terms.  The usefulness of this type of
isolation of curvature parts is evident in general relativity, where it is
consistent with the bundle structure to set the torsion to zero.

\bigskip
\noindent 3.  Flat biconformal space.
\medskip
Before deriving the form of the connection of a flat biconformal space, we
discuss a few of
its properties.
\smallskip
\noindent{{\it {Def:}}} The connection of a flat biconformal space is said
to be in the {\it
{standard flat form}} when it is written in the following way:
$$\eqalignno{
\omega_{0}^{0} &= \alpha_{a}(x) {\bf {d}}x^{a} - y_{a} {\bf {d}}x^{a} \equiv
W_{a} {\bf {d}}x^{a} & (3.1a) \cr
\omega_{0}^{a} &=  {\bf {d}} x^{a} & (3.1b) \cr
\omega_{a}^{0} &=  {\bf {d}} y_{a}  - (\alpha_{a,b} + W_{a}W_{b} -
{\scriptstyle {{1
\over 2}}}W^{2} \eta_{ab}) {\bf {d}}x^{b} &(3.1c) \cr
\omega_{b}^{a} &=   (\eta^{ac} \eta_{bd} - \delta_{d}^{a}
\delta_{b}^{c})W_{c} {\bf
{d}}x^{d} & (3.1d) \cr}$$

\medskip
Notice that the Weyl vector, $W_{a} = \alpha_{a}(x) - y_{a}$, depends on an
arbitrary 4-vector $\alpha_{a}$ and {\it {also}} on the additional four
coordinates $y_{a}$.  The
presence of $\alpha_{a}$ gives the generality required for the
electromagnetic vector
potential, while the $y_{a}$ keeps the dilational curvature zero.  As a
result, unlike Weyl's
theory, the flat biconformal model predicts no size change.  Also notice
that the standard
flat form is preserved by {\it{four-dimensional}} gauge transformations,
$\phi(x)$ and that
the gauge transformation  must be associated with the undetermined vector
field
$\alpha_{a} (x)$.  Thus, the desirable properties of Weyl's original theory
survive in this
more general gauge theory.

The prediction of the exact form of the Weyl vector necessary for
consistently modeling
electromagnetism is nontrivial.  In general, the dilational gauge vector of
biconformal space
is of the form
$$\omega_{0}^{0} = \omega_{0\mu}^{0}(x,y){\bf {d}}x^{\mu} +
\omega_{0}^{0\mu}(x, y) {\bf {d}}y_{\mu}$$
i.e., an 8-dim vector field depending on eight independent variables.
Constraining the
biconformal geometry to have vanishing curvatures forces $\omega_{0}^{0} = (
\alpha_{\mu}(x) - y_{\mu}) {\bf {d}}x^{\mu}$.  This is precisely the form
required to
give the Lorentz force law using the independently established formulation
of Hamiltonian
dynamics in biconformal space.   Different field strengths $\alpha_{[a,b]}$
are in 1-1
correspondence with the possible flat biconformal spaces, so that with the
interpretation of
$\alpha$ as the vector potential, electromagnetic phenomena never lead to
dilations.

We now turn to our first result.
\medskip
\noindent {\it {Thm:}}  When the curvature of biconformal space vanishes,
$\Omega_{B}^{A} = 0$, there exist global coordinates $(x^{a}, y_{a})$ such
that the
connection takes the standard flat form.
\medskip
\noindent {\it {Proof:}} \enskip  Imposing vanishing curvature,
$\Omega_{B}^{A} = 0$, the equations to be solved take the form

\leftskip=.5in
\rightskip=.25in
$$\eqalignno{
{\bf {d\omega}}_{b}^{a} &= \omega_{b}^{c} \wedge \omega_{c}^{a} +
\omega_{b}^{0} \wedge \omega_{0}^{a} - \eta_{bc}\eta^{ad} \omega_{d}^{0}
\wedge
\omega_{0}^{c}  & (3.2a)\qquad \cr
{\bf {d\omega}}_{0}^{a} &= \omega_{0}^{0} \wedge \omega_{0}^{a} +
\omega_{0}^{b} \wedge \omega_{b}^{a}  & (3.2b)\qquad \cr
{\bf {d\omega}}_{a}^{0} &= \omega_{a}^{0} \wedge \omega_{0}^{0} +
\omega_{a}^{b} \wedge \omega_{b}^{0} & (3.2c)\qquad \cr
{\bf {d\omega}}_{0}^{0} &=  \omega_{0}^{a} \wedge \omega_{a}^{0} & (3.2d)
\qquad \cr}$$
The system may be solved by making use of the involution of eq.(3.2b) [38].
This allows
us to consistently set $\omega_{0}^{a} = 0$ and first solve on the subspace
spanned by
the remaining eleven 1-forms.  The initial conditions for these integral
submanifolds
provide a coordinate $x^{a}$ such that $\omega^{a} = {\bf {d}}x^{a}$ with
$x^{a}=const.$ on each leaf.  Each leaf is then a fiber bundle described by
the simpler set of structure equations
$$\eqalignno{
{\bf {d\omega}}_{b}^{a} &= \omega_{b}^{c} \wedge \omega_{c}^{a}  & (3.3a)
\qquad
\cr
{\bf {d\omega}}_{a}^{0} &= \omega_{a}^{0} \wedge {\bf {\omega}}_{0}^{0} +
\omega_{a}^{b} \wedge \omega_{b}^{0} & (3.3b) \qquad \cr
{\bf {d\omega}}_{0}^{0} &= 0 & (3.3c)\qquad \cr}$$
which may be recognized as those of a flat Weyl geometry.  Eq.(3.3c)
implies a pure-gauge
form for the Weyl vector.  Choosing the gauge so that the Weyl vector
vanishes (on the
$\omega_{0}^{a} = 0$  subspace), we are left with the structure equations
for a flat
Riemannian geometry.  Clearly, we have the solution
$$\eqalignno{
\omega_{b}^{a} &= 0   \cr
\omega_{a}^{0} &= {\bf {d}} y_{a}  \cr
\omega_{0}^{0} &= 0 & (3.4) \qquad \cr}$$
Next, we reintroduce the remaining four independent 1-forms
$$\omega_{0}^{a} = {\bf {d}} x^{a} \eqno (3.5) \qquad $$
so that $(x^{a},y_{a})$ provide a set of eight independent coordinates.
From the linearity
of the connection 1-forms in the coordinate differentials,
$\omega_{b}^{a}$, $\omega_{a}^{0}$ and $\omega_{0}^{0}$ will change only by
terms proportional to ${\bf {d}} x^{a}$.  We therefore may write
$$\eqalignno{
\omega_{b}^{a} &= C_{\enskip bc}^{a} {\bf {d}}x^{c}  & (3.6a) \qquad \cr
\omega_{0}^{a} &= {\bf {d}} x^{a} & (3.6b) \qquad \cr
\omega_{a}^{0} &= {\bf {d}} y_{a} +  B_{ab} {\bf {d}}x^{b} & (3.6c) \qquad \cr
\omega_{0}^{0} &= W_{a} {\bf {d}} x^{a} & (3.6d) \qquad \cr}$$
where the coefficients $ W_{a}, B_{ab}$ and $C_{\enskip bc}^{a}$ are
functions of
$x^{a}$ and $y_{a}$ to be found by substitution into the full structure
equations,
eqs.(3.2).  First we determine the form of the Weyl vector, $W_{a}$, from
eq.(3.2d).
Substitution of eqs.(3.6b, c, d) yields
$$\eqalignno{
{\bf {d}}\omega_{0}^{0} &= W_{a}^{\> \>,b}{\bf {d}}y_{b} \wedge {\bf {d}}
x^{a}
+  W_{a,b}{\bf {d}}x^{b} \wedge {\bf {d}} x^{a} \cr
&= {\bf {d}}x^{a} \wedge ( {\bf {d}}y_{a} + B_{ab}{\bf {d}}x^{b}) &(3.7)
\qquad \cr
}$$
Here a lowered comma denotes a partial derivative with respect to $x^{b}$
while a raised
comma denotes a partial with respect to $y_{b}$.  Equating like components
gives
$$\eqalignno{
W_{a}^{\> \>,b} &= - \delta_{a}^{b} &(3.8a) \qquad \cr
B_{[ab]} &= W_{[b,a]} &(3.8b) \qquad \cr}$$
The first of these is immediately integrated to give
$$W_{a} = -y_{a} + \alpha_{a}(x). \eqno(3.9) \qquad $$
Next we substitute eqs.(3.6a, b, d) into eq.(3.2b) to find $C_{\enskip
bc}^{a}$ in terms
of $W_{a}$:
$$\eqalignno{
{\bf {d\omega}}_{0}^{a} &= 0 \cr
&= W_{b} {\bf {d}} x^{b}  \wedge {\bf {d}} x^{a} +  {\bf {d}} x^{b} \wedge
C_{\enskip bc}^{a} {\bf {d}}x^{c}  & (3.10) \qquad \cr}$$
or
$$C_{\enskip [bc]}^{a} + W_{[b} \delta_{c]}^{a} = 0 \eqno(3.11) \qquad  $$
Lowering the upper index in eq.(3.11) and noting that $C_{abc} \equiv
\eta_{ad}
C_{\enskip bc}^{d} = -C_{bac}$, we add two even permutations and subtract
the third to
find
$$C_{\enskip bc}^{a} = - (\delta_{c}^{a} \delta_{b}^{d} - \eta^{ad}
\eta_{bc}) W_{d}
\eqno(3.12) \qquad $$
Finally, $B_{ab}$ is found from eq.(3.2a).  The substitution leads to two
independent
equations, one from ${\bf {d}}x^{a} \wedge {\bf {d}}y_{b}$ cross-terms and
the other
from terms quadratic in ${\bf {d}}x^{a}$.  The first is
$$ (\delta_{c}^{a} \delta_{b}^{d} - \eta^{ad} \eta_{bc}) W_{d}^{\enskip ,
e} =
(\eta^{ae} \eta_{bc} - \delta_{c}^{a} \delta_{b}^{e}) $$
which is identically satisfied by the form of eq.(3.9) for $W_{a}$.
Writing $W^{a}
\equiv \eta^{ab} W_{b}$ and $W^{2} \equiv W^{a}W_{a}$, the second equation
becomes
$$\eqalignno{
\delta^{a}_{d} B_{bc} - \delta^{a}_{c} B_{bd} - B^{a}_{d} \eta_{bc} +
B^{a}_{c}
\eta_{bd}
&= (W^{a}_{\enskip,c}+W^{a}W_{c}) \eta_{bd}-(W^{a}_{\enskip,d}+
W^{a}W_{d}) \eta_{bc} \cr
&\quad - (W_{b,c} + W_{b}W_{c}) \delta^{a}_{d} + (W_{b,d} + W_{b}W_{d})
\delta^{a}_{c}  \cr
&\quad - W^{2}(\delta_{c}^{a} \eta_{bd} - \delta^{a}_{d} \eta_{bc}) &(3.13)
\qquad \cr
} $$
Contraction on the $a$ and $c$ indices gives an expression containing
$B_{ab}$ and its
trace $B = \eta^{ab} B_{ab}$. A second trace yields $B = W^{2} -
W^{a}_{\enskip,a}$.
$B_{ab}$ is then found to be
$$B_{ab} = - (W_{a,b} + W_{a}W_{b} - {\scriptstyle {1 \over 2}}W^{2}
\eta_{ab}) $$
The antisymmetric part of this expression agrees with eq.(3.8b).  The form
for $B_{ab}$
above is now found to solve both eq.(3.13) and the remaining structure
equation,
eq.(3.2c), identically.

We conclude that in flat biconformal space there locally exist coordinates
$x^{a}, y_{a}$
such that
$$\eqalignno{
\omega_{0}^{0} &= \alpha_{a}(x) {\bf{d}}x^{a} - y_{a} {\bf{d}}x^{a}  &(3.1a)
\qquad \cr
\omega_{0}^{a} &=  {\bf{d}}x^{a} &(3.1b) \qquad \cr
\omega_{a}^{0} &=  {\bf{d}}y_{a}  - (\alpha_{a,b} + W_{a}W_{b} - {1 \over
2}W^{2} \eta_{ab}) {\bf{d}}x^{b}  &(3.1c) \qquad \cr
\omega_{b}^{a} &= - (\delta_{d}^{a} \delta_{b}^{c} - \eta^{ac} \eta_{bd})
W_{c}
{\bf{d}}x^{d} & (3.1d) \qquad \cr }$$
That the coordinates $(x^{a}, y_{a})$ are global follows immediately from
the $R^{8}$
topology of the base manifold, completing the proof.

\leftskip=0in
\rightskip=0in

\medskip
We are now in a position to see in more detail how the biconformal
structure corrects
Weyl's theory.  If we hold the $y$-coordinate constant in eqs.(3.1),  we
find that
eqs.(3.1a, b \& d) are the connection forms for a 4-dim Weyl spacetime with
conformally
flat metric $\eta_{ab}$.   The remaining expression, eq.(3.1c), is then
simply a 1-form
constructed from the Weyl-Ricci tensor.  However, the dilational curvature
of a 4-dim
Weyl geometry is given by the curl of the Weyl vector, equivalent here to
the arbitrary curl
of $\alpha_{a}$.  Thus, when viewed from Weyl's 4-dim perspective the
solution gives
unphysical size change.  It is only with the inclusion of the additional
momentum variables
proportional to $y_{a}$ that the dilational curvature can be seen to vanish.

With the interpretation of the additional four dimensions
of biconformal space as momentum variables it is clear that the actual motion
of a particle is eight dimensional.
 Now consider an experiment designed to detect a change in the relative
size of two physical objects, for example, a pair of identical atoms.  In
order to see a change, a comparison must be made before and after moving
the two atoms around some closed spacetime loop.  But such motion
necessarily involves changes in momentum as well so there is necessarily a
change in
$y_{a}$ as well as $x^{a}$, inducing a corresponding loop in flat biconformal
space.  But no loop in flat biconformal space ever encloses nonzero dilational
flux, or ever results in a measurable size change.  The size change computed
in the 4-dim Weyl geometry is seen to be in error because it involves forcing
a closed path while holding the momentum variables constant.  Equivalently,
the error occurs because the expression for the dilational curvature is
incomplete.

We show in the next two sections that there is a direct correspondence
between the
geometric variables of biconformal geometry and the physical phase space
variables of
Hamiltonian dynamics.  Then, in Sec.(6) we apply the Hamiltonian
correspondence to the
general flat biconformal solution.  This leads without further assumption
to the Lorentz law
of force and the identification of $\alpha_{a}$ with the electromagnetic
vector potential.
\medskip

\noindent 4. Hamiltonian dynamics in flat biconformal space
\smallskip
Leaving biconformal geometry for the moment, we turn to a geometric
approach to classical
Hamiltonian dynamics.  We first show that the action of a classical
one-particle system may be used to
generate biconformal spaces.  Then, in Sec.(5), we give a unique
prescription for
generating a {\it {flat}} biconformal space.   Begin with 8-dim extended
phase space with
canonical coordinates $(x^{i}, t; p_{i}, p_{4})$ where $(i, j = 1, 2, 3)$.
We assume we
are given a super-Hamiltonian, ${\cal {H}} = {\cal {H}}(x^{a}, p_{a})$
[43].  Imposing
the constraint, ${\cal {H}} = 0$ then gives $p_{4}$ as a function of the
remaining seven
variables, $p_{4}  = - H(p_{i},x^{i},t)$, with H the usual Hamiltonian.
This constraint
insures that time appears as a parameter rather than an independent
dynamical variable.
Now consider the Hilbert form
$$\omega =  H {\bf {d}} t - p_{i} {\bf {d}} x^{i} \eqno (4.1)$$
The integral of $\omega$ is the action functional.  The exterior derivative
of $\omega$ may
always be factored:
$$\eqalignno{
{\bf {d \omega}} &= {\partial H \over \partial x^{i}} {\bf {d}} x^{i}
\wedge {\bf {d}} t
+ {\partial H \over \partial p_{i}} {\bf {d}} p_{i} \wedge {\bf {d}} t  +
{\bf {d}} x^{i}
\wedge {\bf {d}} p_{i} \cr
&= ( {\bf {d}} x^{i}  -  {\partial H \over \partial p_{i}} {\bf {d}} t )
\wedge ( {\bf {d}}
p_{i}  +  {\partial H \over \partial x^{i}}  {\bf {d}} t ) &(4.2) \cr} $$
Therefore, if we define
$$\eqalignno{
{\bf {\omega}}^{i} &\equiv ({\bf {d}} x^{i} - {\partial H \over \partial
p_{i}} {\bf {d}}
t) \cr
{\bf {\omega}}_{i} &\equiv ({\bf {d}}p_{i} + {\partial H \over \partial
x^{i}} {\bf {d}}
t)   \cr
{\bf {\omega}}^{4} &\equiv {\bf {d}}t \cr
{\bf {\omega}}_{4} &\equiv {\bf {d}}{\cal {H}} = 0 &(4.3) \cr }$$
then we can write
$${\bf {d\omega}} = {\bf {\omega}}^{a} \wedge {\bf {\omega}}_{a}= {\bf
{\omega}}^{i} \wedge {\bf {\omega}}_{i}  \eqno (4.4) $$
This factoring is clearly preserved by local symplectic transformations of
the 6-basis
$(\omega^{i}, \omega_{i})$, as well as reparameterizations of the time.
Obviously these
transformations include the usual canonical transformations of coordinates
as a special
case.  One class of such allowed transformations of basis is achieved by
the addition of
$c_{ab} \omega^{b}$ to ${\bf {\omega}}_{a}$, where $c_{ab} = c_{ba}$.   For
the
moment we take $c_{ab} = 0$, but below we show the existence of a unique
choice of
$c_{ab}$ that leads to a flat biconformal space.

We may also define a connection 1-form, $\omega_{b}^{a}$.  This choice is
uniquely
determined by requiring the resulting biconformal space to be flat, but
first we show that
every possible choice leads to some biconformal space.  Without imposing
flatness any
choice is possible.  For this most general case, let $\omega_{b}^{a}$ be
any linear
combination of $\omega_{0}^{a}$ and $\omega^{0}_{a}$, and define the
(necessarily
horizontal) curvatures to be:
$$\eqalignno{
\Omega_{b}^{a} &\equiv   {\bf {d\omega}}_{b}^{a} - \omega_{b}^{c} \wedge
\omega_{c}^{a} -  \omega_{b}^{0} \wedge \omega_{0}^{a} - \eta_{bc}\eta^{ad}
\omega_{0}^{c} \wedge \omega_{d}^{0} &  (4.5a) \cr
\Omega_{0}^{a} &\equiv {\bf {d\omega}}_{0}^{a} - {\bf {\omega}} \wedge {\bf
{\omega}}_{0}^{a} -  {\bf {\omega}}_{0}^{b} \wedge {\bf {\omega}}_{b}^{a} &
(4.5b) \cr
\Omega_{a}^{0}  &\equiv {\bf {d\omega}}_{a}^{0} - {\bf {\omega}}_{a}^{0}
\wedge
{\bf {\omega}} -  {\bf {\omega}}_{a}^{b} \wedge {\bf {\omega}}_{b}^{0}  &
(4.5c)
\cr
\Omega_{0}^{0} &\equiv {\bf {d\omega}} -  {\bf {\omega}}_{0}^{a} \wedge {\bf
{\omega}}_{a}^{0}  \equiv 0 &  (4.5d) \cr} $$

The necessary presence and form of eq.(4.4) for ${\bf {d\omega}}$, and the
dependence
of the curvatures on {\it {both}} ${\bf {\omega}}^{a}$ and ${\bf
{\omega}}_{a}$ clearly
show this to be a dilationally flat (i.e., $\Omega_{0}^{0} = 0$)
biconformal space.
Therefore, extended phase space together with a Hamiltonian symplectic
structure may be
viewed as a certain kind of biconformal space.  Turning this around, {\it
{we can interpret
biconformal space, and therefore conformal gauge theory, as a
generalization of one-particle phase space.}}

It is worth pointing out that biconformal spaces, while including
Hamiltonian extended
phase spaces as special cases, also contain all 4-dim pseudo-Riemannian
geometries as other special cases.  We therefore have a differential
geometry rich enough to describe both
general relativity and Hamiltonian particle dynamics.  Furthermore, as we
show in Sec.(7), biconformal spaces include the even larger class of all
4-dimensional Weyl geometries, and this class allows the consistent
geometric unification of gravity and electromagnetism.

The presence of both Hamiltonian and Riemannian structures is reassuring,
since it is one aim of the study of biconformal spaces to place relativity
theory and quantum theory in a
common mathematical framework.  While the developments here suggest only
the possibility of describing, perhaps, a quantum particle in a curved
background, the full picture is actually somewhat better because the
symplectic 2-form of the Hamiltonian structure gives a complex structure to
the tangent/co-tangent space and an almost complex structure to the
biconformal space itself.  The biconformal structure therefore gives a
natural complexification of spacetime in such a way that the usual real
structure is immediately evident.  One might hope, for example, to see some
special significance for the Ashtekar connection when biconformal space is
expressed in terms of the $SU(2,2)$ conformal covering group instead of
$O(4,2)$.  Whether these hopes are realized or not is the subject of
current study.

We also note here that the idea of a quantum interpretation of conformal
geometry agrees in some ways with earlier proposals [44-47] relating phase
space, Weyl geometry and quantum physics.  These proposals, however, lack
the full geometric structure of conformal gauge theory, do not demonstrate
the intrinsically biconformal structure of Hamiltonian systems, and use a
different inner product than that proposed in Sec.6 [see also 41].

We next show the relationship of the geometry described by eqs.(4.5) to
classical
mechanics.

Since the curvatures are 2-forms and because ${\bf {d}}{\cal {H}} = 0$,
each of the
curvatures , $\Omega^{a}$ and $\Omega_{a}$ is necessarily at least linear
in one of  the
six basis forms $(\omega^{i}, \omega_{i})$.  Thus, six of the structure
equations,
eqs.(4.5b,c) are in involution.  This involution allows us to set ${\bf
{\omega}}^{i} =
{\bf {\omega}}_{i}  =  0$, thereby singling out a fibration of the bundle
by 1-dim
submanifolds, i.e., the classical paths of motion.  Examination of eq.(4.3)
shows that these
conditions simply give Hamilton's equations of motion:
$$ \eqalignno{  {\bf {d}} x^{i}  &=  {\partial H \over \partial p_{i}} {\bf
{d}} t \cr
{\bf{d}} p_{i}  &=  -  {\partial H \over \partial x^{i}} {\bf {d}} t  &
(4.6) \cr } $$
The Fr\"{o}benius theorem guarantees the existence of solutions to these
equations for the
paths.  The remaining two structure equations then reduce to
$$ {\bf {d\omega}}_{j}^{i} = {\bf {d\omega}} = 0 \eqno (4.7) $$
which are identically satisfied on curves.

It is of interest to further note that on the full bundle the condition
that ${\bf {\omega}}$ be
exact is the Hamiltonian-Jacobi equation, since we may then write ${\bf
{\omega}} = {\bf
{d}} S$.  Substituting for ${\bf {\omega}}$ and expanding ${\bf {d}} S$ gives
$$-H{\bf {d}}t + p_{i} {\bf {d}} x^{i} = {\partial S \over \partial x^{i} }
{\bf {d}}
x^{i} + {\partial S  \over  \partial p_{a} } {\bf {d}} p_{a} + {\partial S
\over \partial t}
{\bf {d}} t \eqno (4.8)$$
so that
$${\partial S \over \partial p_{i} }=0, \quad {\partial S \over \partial
p_{4} }=0 \eqno
(4.9)$$
$${\partial S \over \partial x^{i} }= p_{i} \eqno (4.10)$$
and
$$H({\partial S \over \partial x^{i} },x^{i},t) = -{\partial S \over
\partial t}. \eqno (4.11)
$$
Therefore, since ${\bf {\omega}}$ is the Weyl vector of the generated
biconformal space,
the Hamilton-Jacobi equation holds if and only if the Weyl vector is pure
gauge, ${\bf
{\omega}} = {\bf {d}}S$.  A gauge transformation reduces the Weyl vector to
zero.  Since
the dilational curvature is always zero for the geometry built from a
Hamiltonian system,
when ${\bf {d}}\omega = 0$ we also have
$$\omega^{i} \wedge \omega_{i} = 0, \eqno (4.12)$$
implying three linear dependences between these six solder forms.  Together
with the
vanishing of ${\bf {d}}{\cal {H}}$, the Hamilton-Jacobi equation therefore
specifies a 4-dimensional subspace of the full biconformal space.
\bigskip
\bigskip
\noindent 5.  Flat biconformal space and the Hamiltonian geometry
\smallskip
In this section, we show how to specify a unique {\it {flat}} biconformal
space for a given
Hamiltonian system.  This is achieved by making a judicious choice of the
solder and co-solder forms.

This time we initially regard the super-Hamiltonian as an unconstrained
function of all eight
coordinates.  The Hamiltonian, $H$, is again defined as the solution for
$-p_{4}$ when
${\cal {H}}$ is zero.  Thus, when ${\cal {H}}$ is unconstrained, $p_{4}$ is
also
variable and the Hilbert form generalizes to
$$\omega_{0}^{0} =  -p_{a} {\bf {d}} x^{a} = -p_{4}({\cal {H}},H) {\bf {d}}t -
p_{i} {\bf {d}}x^{i}.  \eqno (5.1)$$
Then
$${\bf {d \omega}_{0}^{0}} = {\bf {d}} x^{a} \wedge {\bf {d}} p_{a}, \eqno
(5.2)$$
where we reserve the symbols $\omega, \omega_{i}$ and $\omega^{i}$ for the
special
case when ${\cal {H}} = 0$ and $p_{4} =  - H(x^{i}, t, p_{i})$.

The solder and co-solder forms $\omega_{0}^{a}$ and $\omega_{a}^{0}$ are now
identified by comparing the expression for ${\bf {d}} \omega_{0}^{0}$ to
the flat
biconformal solution, eqs.(3.1).  To make eq.(3.1a) agree with eq.(5.1) we
must have
$\alpha_{a}(x) = 0$ and $y_{a} = p_{a}$.  We can set $\alpha_{a} = 0$ with
a gauge
change as long as $\alpha_{[a,b]} = 0$.  Then, with $y_{a} = p_{a}$, the
remainder of
the connection is fully determined to be:
$$\eqalignno{
{\bf {\omega}}_{0}^{0} &=  - p_{a} {\bf {d}} x^{a} \cr
{\bf {\omega}}_{0}^{a} &=  {\bf {d}} x^{a} \cr
{\bf {\omega}}_{a}^{0} &=  {\bf {d}} p_{a} - (p_{a}p_{b} - {\scriptstyle {1
\over 2}}
p^{2} \eta_{ab}) {\bf {d}} x^{b}  \cr
{\bf {\omega}}_{b}^{a} &= - (\delta_{d}^{a} \delta_{b}^{c} - \eta^{ac}
\eta_{bd})
p_{c} {\bf {d}} x^{d} & (5.3) \cr  }$$
Note that canonical changes of variable do not change $\omega_{0}^{0}$ by
more than a
scale change, ${\bf {d}}\phi$, so this form of the connection is correct
for any canonical
variables $(x^{i},p_{i}) \longrightarrow (q^{i},\pi_{i})$.

We now restrict to the hypersurface ${\cal {H}} = 0$ so that $ p_{4} = -
H(p_{i},x^{i},t)$.  The symmetric coefficient
$$c_{ab} = p_{a}p_{b} - {\scriptstyle {1 \over 2}} p^{2} \eta_{ab} \eqno
(5.4) $$
provides a symplectic change of basis with respect to the manifestly closed
and
nondegenerate 2-form, ${\bf {d}} \omega_{0}^{0} = \omega_{0}^{a} \wedge
\omega_{a}^{0} = {\bf {d}}x^{a} \wedge {\bf {d}}p_{a}$.  The differential of
$\omega_{0}^{0}$ is still seen to factor as in eq.(4.2)  either directly by
differentiation or
by substitution of $ p_{4} = - H(p_{i},x^{i},t)$ into ${\bf
{\omega}}_{0}^{a} \wedge
{\bf {\omega}}_{a}^{0}$, with $\omega_{0}^{a}$ and $\omega_{a}^{0}$ given by
eqs.(5.3).  The involution of eqs.(4.5b, c) for $\omega_{i}$ and
$\omega^{i}$ still holds,
with the classical curves given by $\omega_{i} = \omega^{i} = 0$.  We
therefore recover
the Hamiltonian system, and have shown it to lie in a unique flat
biconformal space.

The resulting curves in the biconformal space are easily found by first
writing the frame
field in terms of $\omega^{i}$ and $\omega_{i}$:
$$\eqalignno{
\omega_{0}^{0} &= (H - p_{i}{\partial H \over \partial p_{i}}) {\bf {d}}t -
p_{i}
\omega^{i} = L(\dot{x}^{i}, x^{i}, t){\bf {d}}t - p_{i} \omega^{i} \cr
\omega_{0}^{i} &= \omega^{i} +{\partial H \over \partial p_{i}} {\bf {d}}t  \cr
\omega_{0}^{4} &= {\bf {d}}t \cr
\omega_{i}^{0} &= \omega_{i} - (p_{i}p_{j} + {\scriptstyle {1 \over 2}}
(H^{2} -
p^{2}) \delta_{ij}) \omega^{j} + (p_{i}H - {\partial H \over \partial
x^{i}} - {\scriptstyle
{1 \over 2}} (H^{2} - p^{2}) \delta_{ij}{\partial H \over \partial p_{j}})
{\bf {d}} t \cr
\omega_{4}^{0} &= (Hp_{i} - {\partial H \over \partial x^{i}}) \omega^{i} +
(H{\partial
H \over \partial p_{i}} p_{i} -  {\scriptstyle {1 \over 2}} (H^{2} - p^{2})
- {\partial H
\over \partial x^{i}}{\partial H \over \partial p_{i}})  {\bf {d}} t  \cr
\omega_{j}^{i} &= - (\delta_{l}^{i} \delta_{j}^{k} - \delta^{ik} \eta_{jl})
p_{k}
(\omega^{l} + {\partial H \over \partial p_{l}}{\bf {d}}t)  \cr
\omega_{4}^{i} &= H \omega^{i} - (\delta^{ij}p_{j} - H{\partial H \over
\partial p_{i}})
{\bf {d}}t  \cr
\omega_{i}^{4} &= - \delta_{ij} \omega_{4}^{j} & (5.5) \cr}$$
The simple example of a free particle is instructive.  Setting $\omega^{i}
= \omega_{i} =
0$ and $H^{2} = p^{2} + m^{2} \neq 0$,  eqs.(5.5) reduce to:
$$\eqalign{
\omega_{0}^{0} &= {m^{2} \over H} \> {\bf {d}}t = m {\bf {d}}\tau  \cr
\omega_{0}^{a} &= {\eta^{ab} p_{b} \over m^{2} } \> \omega_{0}^{0} = u^{a}{\bf
{d}}\tau  = {\bf {d}}x^{a} \cr
\omega_{a}^{0} &= {\scriptstyle {1 \over 2}}p_{a} \> \omega_{0}^{0} =
{\scriptstyle
{1 \over 2}} m p_{a}{\bf {d}}\tau  \cr
\omega_{b}^{a} &= 0  \cr}$$
where use of  $\omega_{0}^{0}$ or proper time ${\bf {d}}\tau$ in place of
${\bf {d}}t$
simplifies the expressions.  The solder and co-solder forms are
proportional to the
displacement and momentum, respectively.

Also, we see again that the involution required to specify the classical
paths necessarily
exists.  Since the biconformal base space is spanned by the eight forms
${\bf {d}}x^{a}$
and ${\bf {d}}p_{a}$, the only way the involution could fail is if there
was an
independent  ${\bf {d}}t \wedge {\bf {d}}p_{4}$ term in the torsion or
co-torsion of the
$p_{4} = - H$ hypersurface.  But since ${\bf {d}}H$ is given {\it {a
priori}} in terms of
the other seven forms, this cannot happen.
We conclude that the Hamiltonian dynamics of a point particle is equivalent
to the
specification of a hypersurface, $y_{4} = y_{4}(y_{i},x^{a})$, in a flat
biconformal
space, and the consequent necessary existence of preferred congruence of
curves in the
hypersurface.

The principal significance of this result is {\it {the unambiguous
identification of the
geometric quantities which arise in the gauging of the 15-dim conformal
group with
corresponding physical quantities in phase space.  In particular, the extra
four coordinates
are identified with momenta and the integral of the Weyl vector is
identified with the
action}}. This insight should not be construed as a replacement for the
phase space description of particle mechanics.  Rather, the interpretation
presented here is to be regarded as the 1-particle limit of a full
biconformal field theory, and is intended to provide guidance for and a
check on that field theory.  We shall see in Sec.(7) that for curved
biconformal spaces, the ``momentum-like" basis forms $\omega_{a}^{0}$
contain the stress-energy source for the Einstein equation, further
strengthening the present interpretation.

Notice that for multiparticle systems this interpretation of the variables
of flat biconformal space in terms of phase space variables differs from
the usual n-particle phase space of more complicated systems.  In the case
of multiple particles in a small region of (nearly) flat biconformal space,
the particles share {\it {both}} the momentum and configuration space.
Thus, while a single point of a multiparticle phase space characterizes the
entire multiparticle system, the many particle biconformal model will be
described by many points in the same 8-dimensional space. Nonetheless, the
physical interpretation of the extra four coordinates as momenta remains
valid, and each particle treated separately will (locally) obey its own set
of Hamilton's equations.

The same conclusion holds for the fields in biconformal spaces$-$the extra
dimensions will give the local field momentum.  This may be seen in either
of two ways.  First, fields are the continuum limits of multiparticle
systems and their local momentum will therefore be the limit of the
particle momenta.  Second, whenever a single field quantum is confined to
an isolated region which is small relative to the curvature, it will have
an interpretation as a single particle.  The biconformal co-space must then
give the momentum of that particle.

While accomplishing an interpretation of the biconformal variables, we have
also circumvented problems with previous treatments of conformal gauge
theory.  Unlike 4-dimensional conformal gaugings which always reduce to a
4-dimensional Weyl geometry in which the inverse translations are
auxiliary, the current approach retains the full conformal degrees of
freedom.  The extra 4 degrees of freedom are now seen to correspond to the
inclusion of momentum variables in the physical description.
\bigskip
\noindent 6.  Weyl's theory in biconformal space
\medskip
Now consider how the dynamics of the Hamiltonian correspondence of Secs.(4)
and (5) is
modified by the presence of the vector field, $\alpha$.  The Weyl vector is
now given by
$$ \omega_{0}^{0} = -p_{a} {\bf {d}}x^{a} + \alpha_{a}{\bf {d}}x^{a} \equiv -
\pi_{a} {\bf {d}}x^{a} \eqno (6.1) $$
so that
$${\bf {d}} \omega_{0}^{0} = {\bf {d}}x^{a} \wedge {\bf {d}} \pi_{a}  \eqno
(6.2) $$
Specifying $\pi_{4} = \pi_{4} (\pi_{i},x^{a})$ and setting $\omega^{i} =
\omega_{i} =
0$ again leads to Hamilton's equations, in the form
$$\eqalignno{
{\bf {d}} x^{i}  &=  -{\partial \pi_{4}  \over \partial \pi_{i}} {\bf {d}} t & (6.3a) \cr{\bf{d}} p_{i}  &=    {\partial \pi_{4}  \over \partial x^{i}} {\
bf {d}} t & (6.3b) \cr } $$
Maintaining our previous identification $p_{4} = -H = -(m^{2} +
p^{2})^{1/2}$ we find
$$\eqalignno{
 \pi_{4} &= -(m^{2} + p^{2})^{1/2} - \alpha_{4} \cr
&= -(m^{2} + (\pi^{i} + \alpha^{i}) (\pi_{i} + \alpha_{i}))^{1/2} -
\alpha_{4} & (6.4)
\cr }$$
Hamilton's equations become
$$ \eqalign{
{\bf {d}} x^{i} &=  {\pi^{i} + \alpha^{i}  \over (m^{2} + (\pi +
\alpha)^{2})^{1/2}}
{\bf {d}} t \cr
&=  {p^{i}  \over (m^{2} + p^{2})^{1/2}} {\bf {d}} t \cr }$$
or  $$  {\dot{x}}^{i} =  {p^{i}  \over (m^{2} + p^{2})^{1/2}}  \eqno(6.5) $$
for the position variables and
$$\eqalignno{
{\bf {d}} \pi_{i}  &=  - {\partial \alpha_{4} \over \partial x^{i} }
\enskip {\bf {d}} t  -
{\pi^{j} + \alpha^{j}  \over (m^{2} + (\pi + \alpha)^{2})^{1/2}} \enskip
\alpha_{j,i}
{\bf {d}} t \cr
&=  -{\partial \alpha_{4} \over \partial x^{i} } {\bf {d}} t  -
{\dot{x}}^{j} \alpha_{j,i}
{\bf {d}} t &(6.6) \cr }$$
for the momentum.  The left hand side of eq.(6.6) expands to
$$ {\bf {d}} \pi_{i} = {\bf {d}}p_{i} -  { \partial \alpha_{i} \over
\partial t} {\bf {d}}t -
\alpha_{i,j} {\bf {d}}x^{j} $$
which finally leads to
$$ {dp_{i} \over dt} = (-{\partial \alpha_{4} \over \partial x^{i} } +  {
\partial \alpha_{i}
\over \partial t}) + {\dot{x}}^{j}(\alpha_{i,j} - \alpha_{j,i}) \eqno(6.7a) $$
For the time component we have
$$\eqalignno{
{dp_{4} \over dt } &= {d \over dt} (\pi_{4} + \alpha_{4}) = {d \over dt}
(-(m^{2} +
p^{2})^{1/2})  \cr
&=- {1 \over (m^{2} + p^{2})^{1/2} } p^{i} {dp_{i} \over dt} \cr
&= - {\dot{x}}^{i} {dp_{i} \over dt} \cr
&=  {\dot{x}}^{i}  ({\partial \alpha_{4} \over \partial x^{i} } -  {
\partial \alpha_{i} \over
\partial t}) &(6.7b) \cr }$$
Eqs.(6.7) give the Lorentz force law if we identify
$$ \alpha_{a} = q(\phi, -A_{i}) = - qA_{a}$$
Thus, the existence and form of the electromagnetic force on a charged
particle is correctly
predicted by the general solution for flat biconformal space.  The presence
of the vector
field $\alpha_{a}$, its gauge dependence and its proper coupling to matter
are direct
consequences of the local gauge theory of scalings and of our identificaton
of Hamiltonian
dynamics as the specification of a hypersurface in flat biconformal space.
Nor does this
formulation suffer the objection made to the original Weyl theory.  Since
we are studying
precisely the flat biconformal spaces, there is no dilational curvature and
no measurable size
change.

In concluding this section, we comment briefly on the relationship of these
considerations
to the standard U(1) model of electromagnetic field theory.  Biconformal
space has a
natural metric structure based on the Killing metric of the underlying
conformal group.  The
Killing metric has eigenvalues $\pm 1$ and zero signature, hence the local form
$$K_{AB} = diag \left( \eta_{ab}, -\eta^{ab} \right) \eqno(6.8)$$
where $(A, B) = (1, \dots, 8)$.  Consequently, the proportionality constant
in the
identification between $y_{a}$ and $p_{a}$ is purely imaginary.  Since the
vector field
$\alpha_{a}(x)$ is real, we have
$$\eqalignno{
y_{a} &= \lambda(ip_{a} + \alpha_{a}) &(6.9a) \cr
&= i\lambda(p_{a} - i  \alpha_{a}) &(6.9b) \cr } $$
as the relationship between geometric and physical variables.  Since the
proportionality
constant drops out of Hamilton's equations it is not measurable
classically, but it must be
included whenever particle paths are allowed to deviate from the classical
trajectories.
Thus, the real-valued 4-dim scale invariance that preserves the form of the
standard flat
biconformal connection becomes a U(1) invariance when applied to the
physical variables
of eq.(6.9b).
\bigskip
\noindent 7.  Coupling to gravity and other fields
\medskip
So far, we have motivated a new 8-dimensional gauging of the conformal
group, found the class of flat solutions of the resulting biconformal
spaces and showed how to interpret the flat solutions as a phase space for
a single particle coupled to a background electromagnetic field.  In this
section, we generalize these results to gravitating biconformal spaces in
which the solder form satisfies the Einstein equation with arbitrary matter
as source, and the vector potential identified in the previous sections
satisfies the Maxwell field equations with arbitrary electromagnetic
currents.  The results of this section therefore comprise a unified
geometric theory of gravitation and electromagnetism.

Our starting point is the full set of structure equations, eqs.(2.12),
which define the fifteen curvature 2-forms $\Omega_{B}^{A}$.  Each of these
curvatures has the three biconformally invariant terms displayed in
eq.(2.13).  In addition, we note that the two-form ${\bf
{d}}\omega_{0}^{0}$ is separately biconformally invariant, and the solder
and co-solder forms $\omega_{0}^{a}$ and $\omega_{a}^{0}$ transform
tensorially.  As long as the corresponding Bianchi identities are
satisfied, a specification of any combination of these fields may be used
to invariantly determine sub-classes of biconformal geometries.

To begin, we seek some general constraints to limit the number of
independent fields.  For classical geometries it is reasonable to assume
that no classical path in phase space encloses a plaquette on which the
dilation, $\Omega_{0}^{0}$ is nonvanishing.  The simplest (but by no means
the only) way to guarantee this is to just set $\Omega_{0}^{0} = 0$.  We
also expect that the spacetime torsion will vanish in typical classical
models, and again make the simplest hypothesis, that the full torsion (but
{\it {not}} the co-torsion) is zero.  Thus we have the general constraints
$$\Omega_{0}^{0} = 0 \eqno (7.1)$$
$$\Omega_{0}^{a} = 0 \eqno (7.2)$$
Next, we note that the vanishing of the torsion puts the solder form in
involution.  Assuming the resulting foliation to be regular, there exists a
4-dimensional sub-manifold of the base space spanned by $\omega_{0}^{a} =
{\bf {e}}^{a}$.  As a final general constraint, we require the existence of
a completion ${\bf {f}}_{a}$ to the ${\bf {e}}^{a}$ basis in which the
spacetime curvature is traceless:
$$\Omega_{bac}^{a} = 0  \hskip.25in (in\>the\>({\bf {e}}^{a}, {\bf
{f}}_{a})\>basis) \eqno(7.3)$$
We shall show that these three constraints are sufficient for the resulting
class of geometries to take a recognizable form.  There is {\it {no}}
restriction of the dependence of any of the fields on the eight
coordinates, eg. $(x^{\mu}, y_{\nu})$.

In addition to the general constraints above, we posit two field equations.
The first provides a source for the Weyl vector, which we take here to be
in the typical form for an electromagnetic current
$$*{\bf {d}}*{\bf {d}}\omega_{0}^{0} = {\bf {J}} = J_{a}(x) {\bf {e}}^{a}
\eqno(7.4a) $$
Finally, part of the co-solder form is determined by an arbitrary
stress-energy tensor $T_{ab}$ via
$$\omega_{a}^{0} = {\cal {T}}_{a} + \ldots  \eqno(7.4b) $$
where ${\cal {T}}_{a} \equiv  - {1 \over 2}(T_{ab}- {1 \over 3} \eta_{ab}T)
{\bf {e}}^{b}$.  The consistency of these expressions and the form of the
remaining part of the co-solder form are established using the general
equations, eqs.(7.1)-(7.3).

The central result of this section is that a biconformal space is in 1-1
correspondence with a 4-dimensional Einstein-Maxwell spacetime if and only
if eqs.(7.1)-(7.4) hold.  The 1-1 correspondence is based on an isomorphism
between the biconformal co-space at $x^{\mu}_{0}$ and the tangent space to
a 4-dimensional spacetime at a corresponding point $x^{\mu}_{0}$.  A more
complete presentation of the techniques used and some related results are
given in [51].

The equations to be solved are
$$\eqalignno{
{\bf {d\omega}}_{b}^{a} &= \omega_{b}^{c} \wedge \omega_{c}^{a} +
\omega_{b}^{0} \wedge \omega_{0}^{a} - \eta_{bc}\eta^{ad} \omega_{d}^{0}
\wedge
\omega_{0}^{c} + \Omega_{b}^{a} &(2.12a) \cr
{\bf {d\omega}}_{0}^{a} &= \omega_{0}^{0} \wedge \omega_{0}^{a} +
\omega_{0}^{b} \wedge \omega_{b}^{a} &(2.12b') \cr
{\bf {d\omega}}_{a}^{0} &= \omega_{a}^{0} \wedge \omega_{0}^{0} +
\omega_{a}^{b} \wedge \omega_{b}^{0} + \Omega_{a}^{0} &(2.12c) \cr
{\bf {d\omega}}_{0}^{0} &= \omega_{0}^{a} \wedge \omega_{a}^{0}  &
(2.12d') \cr} $$
The first part of our proof follows Theorem I of [51].  We begin with the
Bianchi identity for eq.(2.12b$'$), which follows by taking the exterior
derivative and using ${\bf {d}}^{2} = 0$.  The result is
$$ \omega_{0}^{b} \wedge \Omega_{b}^{a} = 0 \eqno(7.5)$$
which in particular shows that the momentum term of the curvature vanishes
$$ \Omega_{b}^{acd} = 0 \eqno(7.6) $$
Similarly, the Bianchi identity for eq.(2.12d$'$) requires
$$ \Omega_{b}^{0cd} = 0 \eqno(7.7) $$
for the co-torsion.  Together with the vanishing of the dilation and the
torsion, eqs.(7.6) and (7.7) imply that the bundle is momentum flat, i.e.,
the momentum term of each curvature vanishes.

Next, we use the involution of the solder form, as noted above.  The
involution means that the biconformal bundle is foliated by 11-dimensional
sub-bundles on which $\omega_{0}^{a} = 0$.  On this sub bundle, using the
momentum-flatness, the structure equations reduce to
$$\eqalignno{
{\bf {d\omega}}_{b}^{a} &= \omega_{b}^{c} \wedge \omega_{c}^{a}  &(7.8a) \cr
{\bf {d\omega}}_{a}^{0} &= \omega_{a}^{0} \wedge \omega_{0}^{0} +
\omega_{a}^{b} \wedge \omega_{b}^{0} &(7.8b) \cr
{\bf {d\omega}}_{0}^{0} &= 0 &(7.8c) \cr} $$
Eqs.(7.8) are just the structure equations for a flat 4-dimensional Weyl
geometry.  Eq.(7.8a) shows that we can perform a Lorentz gauge
transformation on the entire bundle such that
$\omega_{b}^{a}\vert_{\omega_{0}^{a}=0} = 0$, while eq.(7.8c) shows the
existence of a scaling such that $\omega_{0}^{0}\vert_{\omega_{0}^{a}=0} =
0$.  Following these gauge changes eq.(7.8b) takes the simple form
$${\bf {d\omega}}_{a}^{0} = 0  $$
with the immediate solution
$$\omega_{a}^{0}\vert_{\omega_{0}^{a}=0} = f_{a}^{\enskip \mu}(x) {\bf
{d}}y_{\mu} \eqno(7.9)$$
where we have introduced four coordinates $y_{\mu}$ to span the
4-dimensional base space of the sub-manifold.  Notice that the coefficient
matrix $ f_{a}^{\enskip \mu}(x)$ must be nondegenerate.

Now, assuming the foliation to be regular, the space of leaves of the
foliation is also a 4-dimensional manifold spanned by $\omega_{0}^{a}$.
There exist coordinates $x^{\mu}$ such that
$$\omega_{0}^{a} = e_{\mu}^{\enskip a}(x, y) {\bf {d}}x^{\mu} \eqno(7.10) $$
so that each leaf of the foliation is given by some constant value $x^{\mu}
= x^{\mu}_{0}$.

From the sub-bundle, with connection $\omega_{b}^{a}\vert_{x_{0}} = 0$,
$\omega_{0}^{0}\vert_{x_{0}} = 0$ and $\omega_{a}^{0}\vert_{x_{0}} = {\bf
{f}}_{a}$, we extend back to the full bundle by allowing $x$ to vary.  This
can change each connection form at most by a term proportional to ${\bf
{d}}x^{\mu}$ or equivalently ${\bf {e}}^{a}$, so the full connection may
always be given the local form
$$\eqalignno{
\omega_{b}^{a} &= C_{bc}^{a} {\bf {e}}^{c} &(7.11a) \cr
\omega_{0}^{a} &= {\bf {e}}^{a}(x, y) &(7.11b)  \cr
\omega_{a}^{0} &= f_{a}^{\enskip \mu}(x) {\bf {d}}y_{\mu}  + B_{ab}{\bf
{e}}^{b} \equiv {\bf {f}}_{a} + {\bf {B}}_{a} &(7.11c) \cr
\omega_{0}^{0} &= W_{a}{\bf {e}}^{a}  &(7.11d) \cr }$$
where $W_{a}, B_{ab}, $ and $C_{bc}^{a}$ are functions of all 8 coordinates
$(x^{\mu}, y_{\nu})$.  The functional dependence of $e_{\mu}^{\enskip a}$,
while {\it {a priori}} unspecified, is actually restricted by
eq.(2.12b$'$).  Since $\omega_{b}^{a}$ and $\omega_{0}^{0}$ depend only on
${\bf {e}}^{a}$ and not ${\bf {f}}_{a}$, writing out ${\bf {de}}^{a}$ in
coordinates immediately shows that $\partial^{\nu} e_{\mu}^{\enskip a} =
0$, where we write $\partial^{\nu} \equiv {\partial \over \partial y_{\nu}
}$ and $\partial_{\nu} \equiv {\partial \over \partial x^{\nu}}$.
Therefore, we may drop the $y$-dependence in eq.(7.11b).

If we had made a different initial choice of the coordinates $y_{\mu}$, the
connection would still be of the form given by eqs.(7.11), although $W_{a},
B_{ab}, $ and $C_{bc}^{a}$ would be different functions.  We can therefore
assume without loss of generality that the basis $({\bf {e}}^{a}, {\bf
{f}}_{b})$ has been chosen as the one in which the spacetime curvature is
tracefree.

Next, after dividing $B_{ab}$ into symmetric and antisymmetric parts
$$B_{ab} \equiv S_{ab}(x) + F_{ab}(x) \eqno(7.12)$$
with $S_{ab}= S_{(ab)}$ and $F_{ab}= F_{[ab]}$, we substitute eq.(7.11c)
into the dilation equation, eq.(2.12d$'$), to find
$$\eqalignno{
{\bf {d\omega}}_{0}^{0} &= {\bf {e}}^{a} \wedge {\bf {f}}_{a} + {\bf {F}} \cr
&= {\bf {d}}x^{\mu} \wedge f_{\mu}^{\enskip \nu}(x) {\bf {d}}y_{\nu} + {\bf
{F}} &(7.13) \cr } $$
where ${\bf {F}} \equiv F_{ab}{\bf {e}}^{a} \wedge {\bf {e}}^{b} = F_{\mu
\nu}{\bf
{d}}x^{\mu} \wedge {\bf {d}}x^{\nu}$.  The mixed terms of eq.(7.13) yield
$$W_{\mu}^{\enskip ,\nu} = - f_{\mu}^{\enskip \nu}(x) \eqno(7.14)$$
which is immediately integrated to give
$$\omega_{0}^{0}\equiv W_{\mu} {\bf
{d}}x^{\mu}= (-f_{\mu}^{\enskip \nu}(x) y_{\nu} + A_{\mu}(x)){\bf
{d}}x^{\mu}  \eqno(7.15)$$
up to a gauge transformation.  Here $A_{\mu}(x)$ is an arbitrary
integration constant for the $y$-integration.

We could continue to carry the field $f_{\mu}^{\enskip \nu}(x)$ through the
remainder of the general constraints, but it is simpler to employ the field
equation to remove it now.  In a more general class of models,
$f_{\mu}^{\enskip \nu}(x)$ provides an additional unspecified set of
fields.  While it is easy to conjecture that these extra fields may lead to
a geometric electroweak theory, $f_{\mu}^{\enskip \nu}(x)$ appears on the
surface to be a translational rather than a rotational Yang-Mills field, so
that the required quadratic terms are missing from the field strength.  The
proper role of these fields is under study.  For our present purpose, we
simply note that the field equation
$$*{\bf {d}}*{\bf {d}}\omega_{0}^{0} = {\bf {J}} = J_{a}(x) {\bf {e}}^{a}
\eqno(7.3) $$
has no source for the part of $\omega_{0}^{0}$ linear in $y$, and therefore
gives
$$[*{\bf {d}}*{\bf {d}} f_{\mu}^{\enskip \nu}(x)] y_{\nu} = 0 \eqno(7.16) $$
for the $y$-dependent part of the spacetime term.  This is a 4-dimensional
electromagnetic-type equation for the ``potential" $f_{\mu} \equiv
f_{\mu}^{\enskip \nu}(x) y_{\nu}$, which has the unique solution
$f_{\mu}^{\enskip \nu}(x) = 0$ for vanishing boundary conditions at
infinity.  To this solution we add the particular solution $
f_{\mu}^{\enskip \nu} = const.$, since $ f_{\mu}^{\enskip \nu}(x)$ must be
invertible as noted above.  Finally, a constant change of the
$y$-coordinates gives $ f_{\mu}^{\enskip \nu} = \delta_{\mu}^{\nu}$,
simplifying the co-solder form to
$$\omega_{a}^{0} = e^{\enskip \mu}_{a}(x) {\bf {d}}y_{\mu} + B_{a
\mu}(x,y){\bf
{d}}x^{\mu}  \eqno(7.17) $$
where $e^{\enskip \mu}_{a}$ is the inverse to $ e_{\mu}^{\enskip a}$.

With the solder form given by eq.(7.17), eq.(7.15) becomes
$$\omega_{0}^{0}\equiv W_{\mu} {\bf
{d}}x^{\mu}= (- y_{\mu} + A_{\mu}(x)){\bf {d}}x^{\mu}  \eqno(7.18)$$
Thus, this entire class of geometries has the same ``minimal coupling" form
of the Weyl vector as we found for the flat case.  Notice also that the
form of eq.(7.18) is not affected by a purely $x$-dependent gauge
transformation, which is consistent with the interpretation of ${\bf {A}}$
as the electromagnetic vector potential.

Next, we move to eq.(2.12b$'$), which becomes
$${\bf {de}}^{a} = \omega_{0}^{0} \wedge {\bf {e}}^{a} + {\bf {e}}^{b} \wedge
\omega_{b}^{a} \eqno(7.19) $$
This may be uniquely solved for the connection $\omega_{b}^{a}$.  Let
$$\omega_{b}^{a} = \alpha_{b}^{a} + \beta_{b}^{a} \eqno(7.20)$$
where $\alpha_{b}^{a}$ is the usual metric compatible spin connection
satisfying
$${\bf {de}}^{a} = {\bf {e}}^{b} \wedge \alpha_{b}^{a} \eqno(7.21) $$
and require
$${\bf {e}}^{b} \wedge \beta_{b}^{a}+ \omega_{0}^{0} \wedge {\bf {e}}^{a} = 0
\eqno(7.22)$$
Eq.(7.22) is solved uniquely by
$$ \beta_{b}^{a} = - W_{b} {\bf {e}}^{a} + \eta_{bc}\eta^{ad} W_{d} {\bf
{e}}^{c}
\eqno(7.23)$$
so we have now satisfied two of the four structure equations.

Next, we impose the tracelessness condition for the spacetime term of the
curvature in the $({\bf {e}}^{a}, {\bf {f}}_{b})$ basis:
$$\Omega_{\enskip bac}^{a} = 0 \eqno(7.24)$$
Rearranging eq.(2.12a) to solve for $\Omega_{b}^{a}$ we find
$$\eqalignno{
\Omega_{b}^{a} &= {\bf {d\omega}}_{b}^{a} - \omega_{b}^{c} \wedge
\omega_{c}^{a} - \omega_{b}^{0} \wedge \omega_{0}^{a} + \eta_{bc}\eta^{ad}
\omega_{d}^{0} \wedge \omega_{0}^{c}  &(2.12a) \cr
&= ({\bf {d}}\alpha_{b}^{a} - \alpha_{b}^{c} \wedge \alpha_{c}^{a}) +  ({\bf
{d}}\beta_{b}^{a} - \beta_{b}^{c} \wedge \alpha_{c}^{a} - \alpha_{b}^{c}
\wedge
\beta_{c}^{a}) - \beta_{b}^{c} \wedge \beta_{c}^{a}
- \omega_{b}^{0} \wedge {\bf {e}}^{a} + \eta_{bc}\eta^{ad} \omega_{d}^{0}
\wedge
{\bf {e}}^{c}\cr
&\equiv {\bf {R}} _{b}^{a}(\alpha)  +  {\bf {D}}\beta_{b}^{a} -
\beta_{b}^{c} \wedge
\beta_{c}^{a} - \omega_{b}^{0} \wedge {\bf {e}}^{a} + \eta_{bc}\eta^{ad}
\omega_{d}^{0} \wedge {\bf {e}}^{c} &(7.25) \cr }$$
where we have written ${\bf {D}}$ for the ${\bf {e}}^{a}$-compatible
covariant exterior derivative, using the connection $\alpha_{b}^{a}$, and
${\bf {R}} _{b}^{a} \equiv {\bf {d}}\alpha_{b}^{a} - \alpha_{b}^{c} \wedge
\alpha_{c}^{a}$ is the usual Riemann curvature 2-form.  Substituting
eq.(7.17) for $\omega_{a}^{0}$ and and (7.23) for $\beta_{b}^{a}$, some
algebra leads to
$$\eqalignno{
\Omega_{b}^{a} =  &{\bf {R}} _{b}^{a}  +  \eta_{bc}[{\bf {D}}W^{c} {\bf
{e}}^{a}
- {\bf {D}}W^{a} {\bf {e}}^{c} + W^{2}{\bf {e}}^{c}{\bf {e}}^{a}] \cr
&+ {\bf {f}}_{b} {\bf {e}}^{a} - \eta_{bc} \eta^{ad}{\bf {f}}_{d} {\bf
{e}}^{c} +
{\bf {B}}_{b} {\bf {e}}^{a} - \eta_{bc} \eta^{ad}{\bf {B}}_{d}{\bf {e}}^{c}
&(7.26) \cr }$$
Finally, expand $${\bf {D}}W^{a} = -\eta^{ab}{\bf {D}}y _{b} + {\bf {D}}A^{a}
\eqno(7.27) $$
While ${\bf {D}}A^{a}$ is independent of $y_{\mu}$, we must further expand
$$\eqalignno{
{\bf {D}}y_{b} &\equiv {\bf {d}}(e_{b}^{\enskip \mu}y_{\mu}) -
\alpha_{b}^{a} e_{a}^{\enskip \mu} y_{\mu} \cr
&= e_{b}^{\enskip \mu}{\bf {d}}y_{\mu} + ({\bf {d}}e_{b}^{\enskip \mu}-
\alpha_{b}^{a} e_{a}^{\enskip \mu}) y_{\mu} &(7.28)  \cr} $$
The first term in eq.(7.28) is simply ${\bf {f}}_{a}$.  For the final term
we use the Christoffel connection $\Gamma^{\mu}_{\enskip \alpha \beta}$ and
the covariant constancy of $e_{b}^{\enskip \mu}$,
$${\bf {d}}x^{\beta} D_{\beta} e_{b}^{\enskip \mu} = {\bf
{d}}e_{b}^{\enskip \mu} - \alpha_{b}^{a} e_{a}^{\enskip \mu} +
e_{b}^{\enskip \alpha}\Gamma^{\mu}_{\enskip \alpha \beta}{\bf {d}}x^{\beta}
= 0 \eqno(7.29) $$
to write
$$\eqalignno{
{\bf {D}}y_{b} &= {\bf {f}}_{b} - e_{b}^{\enskip
\alpha}\Gamma^{\mu}_{\enskip \alpha \beta}{\bf {d}}x^{\beta} y_{\mu} \cr
&\equiv {\bf {f}}_{b} - \Gamma_{bc} {\bf {e}}^{c} \cr
&\equiv {\bf {f}}_{b} - \Gamma_{b} &(7.30) \cr }$$
Now substitution into eq.(7.26) shows that the cross- and momentum-terms of
the curvature vanish while the components of the spacetime-term become
$$\eqalignno{
\Omega_{bcd}^{a} =  &{\bf {R}} _{bcd}^{a} - (D_{c}A_{b} + \Gamma_{bc} +
B_{bc}) \delta_{d}^{a} + (D_{d}A_{b} + \Gamma_{bd} + B_{bd}) \delta_{c}^{a}
\cr
&+ \eta_{bf} \eta^{ag} (D_{c}A_{g} + \Gamma_{gc} + B_{gc}) \delta_{d}^{f} -
\eta_{bf} \eta^{ag}(D_{d}A_{g} + \Gamma_{gd} + B_{gd}) \delta_{c}^{f} \cr
&-W_{c}(W_{b}\delta^{a}_{d} - \eta_{bf} \eta^{ag}W_{g}\delta_{d}^{f}) \cr
&+W_{d}(W_{b}\delta^{a}_{c} - \eta_{bf} \eta^{ag}W_{g}\delta_{c}^{f}) \cr
&+ W^{2}(\eta_{bc}\delta^{a}_{d} -  \eta_{bd}\delta^{a}_{c})  &(7.31) \cr }$$
Next, by setting the trace of $\Omega_{bcd}^{a}$ to zero, we can insure
that the Einstein
equation holds with arbitrary stress-energy tensor.  Contracting, we find
that the
antisymmetric part $\Omega_{acb}^{c}-\Omega_{bca}^{c}$ is identically zero
while
$$\eqalignno{
\Omega_{acb}^{c} &= R_{ab} + 2(A_{(a;b)} + \Gamma_{ab} + S_{ab} +
W_{a}W_{b}) + \eta_{ab} (A_{c}^{\enskip ;c} + \Gamma_{c}^{\enskip c} +
S_{c}^{\enskip c} - 2W^{2}) = 0 &(7.32) \cr }$$
relates $B_{(ab)} = S_{ab}$ to the Ricci tensor and the vector potential.
If we require
$$\eqalignno{
S_{ab} &\equiv {\cal {T}}_{ab} - (A_{(a;b))} +  \Gamma_{ab} + W_{a}W_{b} -
{1 \over 2} \eta_{ab}W^{2}) &(7.33) \cr }$$
where ${\cal {T}}_{ab} \equiv  - {1 \over 2}(T_{ab}- {1 \over 3}
\eta_{ab}T)$ and
$T_{ab}$ is the electromagnetic stress-energy tensor constructed from $A_{b}$
plus the stress-energy tensor from whatever other phenomenological fields
one wishes to
add, then eq.(7.32) reduces to the Einstein equation with source $T_{ab}$.
We shall see below that the choice given by eq.(7.33) for $S_{ab}$ greatly
simplifies the expressions for both the curvature and the co-torsion.

Nothing further is required in order to satisfy the final structure
equation, eq.(2.12c).
Instead, we {\it {define}} the co-torsion by eq.(2.12c), giving
$$\eqalignno{
\Omega_{a}^{0} &= {\bf {d\omega}}_{a}^{0} - \omega_{a}^{b} \wedge
\omega_{b}^{0} - \omega_{a}^{0} \wedge \omega_{0}^{0}  &(2.12c) \cr } $$
This is easiest to evaluate if we write $\omega_{a}^{0}$ in terms of
$W_{a}$ wherever
possible, resulting in
$$\omega_{a}^{0} = {\cal {T}}_{a} - {\bf {D}}W_{a} - W_{a}{\bf {W}} + {1 \over
2}W^{2}\eta_{ab}{\bf {e}}^{b}  \eqno(7.34) $$
Using the tracefree condition (i.e., the Einstein equation) to replace
${\cal {T}}_{a}$ by ${\cal {R}}_{a} \equiv - {1 \over 2}(R_{ab} - {1 \over
6}\eta_{ab}R) {\bf {e}}^{b}$ we find that eq.(2.12c) takes the form
$$\eqalignno{
\Omega_{a}^{0} &= {\bf {D }}\omega _{a}^{0} - \beta_{a}^{b} \wedge
\omega_{b}^{0} - \omega_{a}^{0} \wedge {\bf {W}}   \cr
&= {\bf {D }}\omega _{a}^{0} + W_{a}{\bf {dW}}- W^{b}\eta_{ac}{\bf {e}}^{c}
\omega_{b}^{0} -  \omega_{a}^{0} \wedge {\bf {W}}  &(7.35) \cr } $$
Substituting eq.(7.34) for $\omega_{a}^{0}$ and using the Ricci identity
${\bf {D}}^{2}\omega_{a} = -{\bf {R}}^{b}_{a} \omega_{b}$ (for an arbitrary
1-form $\omega_{a}$) gives, after several
cancellations, the surprisingly simple result
$$\eqalignno{
\Omega_{a}^{0} &= {\bf {D}}{\cal {R}}_{a} + {\bf {R}}^{b}_{a} -
(\delta_{a}^{c} \delta_{d}^{b} - \eta^{bc} \eta_{ad}){\cal {R}}_{c}{\bf
{e}}^{d} W_{b} \cr
&= {\bf {D}}{\cal {R}}_{a} + {\bf {C}}^{b}_{a}  W_{b} &(7.36) \cr }$$
where ${\bf {C}}^{b}_{a}$ is the Weyl curvature 2-form.  Notice that all
derivatives of
the Weyl vector have cancelled, so the resulting co-torsion also has
vanishing momentum-
and cross-terms.  If we use the contracted second Bianchi identity to write
${\bf {D}}{\cal {R}}_{a}$ as a divergence of the Weyl curvature, we find
$$\eqalignno{
\Omega_{a}^{0} &= - D_{b}{\bf {C}}^{b}_{a} + {\bf {C}}^{b}_{a}  W_{b}
&(7.37) \cr }$$
so the co-torsion is simply the Weyl-covariant (not
$\alpha_{b}^{a}$-covariant) divergence of the Weyl curvature tensor, hence
a direct measure of the deviation of the underlying spacetime from
conformally flat [42].

Finally, remarkable cancellations also occur if we use eq.(7.33) to replace
${\bf {B}}_{a}$ in the full curvature tensor given by eq.(7.26).  The
result is simply
$$\eqalignno{
\Omega_{b}^{a} &= {\bf {C}}^{b}_{a}. &(7.38) \cr }$$

The biconformal geometry is now fully specified except for the vector
potential $A_{\mu}$.  This is fixed by the field equation
$$*{\bf {d}}*{\bf {d}}\omega_{0}^{0} = *{\bf {d}}*{\bf {d}}{\bf {A}} = {\bf
{J}} \eqno(7.4a) $$
where ${\bf {J}}$ is the electromagnetic current.

To complete the proof of the sufficiency of eqs.(7.1)-(7.4) for the
biconformal space to correspond to an Einstein-Maxwell spacetime we only
need to show that the biconformal geometry is effectively 4-dimensional.
This happens because the extra dimensions of the biconformal base space can
be identified with the tangent space of the Riemannian spacetime.  Such an
identification works for two reasons.  First, the co-vector $y_{\mu}$ has
the same scale and Lorentz transformation properties as the basis vectors
${\partial \over \partial x^{\mu} }$ for the tangent space, so we can
identify the bases.  Second, since both the co-space and the tangent space
may be taken as Minkowski vector spaces (i.e., both are flat), the two
complete vector spaces may be identified.  As discussed further in the
final section, this fact guarantees that the extra 4-dimensions do not lead
to undesired new macroscopic effects.

Notice that, while the content of the biconformal space does not exceed
that of a Einstein-Maxwell spacetime, we cannot claim that a biconformal
space satisfying eqs.(7.1)-(7.4) is homeomophic to a 4-dimensional Weyl geometry because the Weyl vector of the biconformal space retains linear $y_{\mu}$ d
ependence.  It is this difference that leads to the vanishing dilation of
the biconformal space.

The converse, that a 4-dimensional Einstein-Maxwell spacetime extends to a
unique biconformal space satisfying eqs.(7.1)-(7.4) is immediate, since,
given the stress-energy tensor and electromagnetic current determining the
spacetime, we can find the solder form and the electromagnetic vector
potential and from these directly write down the connection of the
associated biconformal space as
$$\eqalignno{
\omega_{0}^{0} &= (-y_{\mu} + A_{\mu}){\bf {d}}x^{\mu}  &(7.39a) \cr
\omega_{0}^{a} &= {\bf {e}}^{a}(x) &(7.39b)  \cr
\omega_{a}^{0} &= {\cal {T}}_{a} - {\bf {D}}W_{a} - W_{a}{\bf {W}} + {1 \over
2}W^{2}\eta_{ab}{\bf {e}}^{b} &(7.39c) \cr
\omega_{b}^{a} &= \alpha_{b}^{a} - W_{b} {\bf {e}}^{a} + \eta_{bc}\eta^{ad}
W_{d} {\bf {e}}^{c} &(7.39d) \cr }$$
where we invert the vector space isomorphism above to map the tangent space
into the extra 4-dimensions of the biconformal space.  The form above for
the connection leads directly to the biconformal curvatures
$$\eqalignno{
\Omega_{0}^{0} &=  0  &(7.40a) \cr
\Omega_{0}^{a} &= 0 &(7.40b) \cr
\Omega_{a}^{0} &= - D_{b}{\bf {C}}^{b}_{a} + {\bf {C}}^{b}_{a}  W_{b}
&(7.40c) \cr
\Omega_{b}^{a} &= {\bf {C}} _{b}^{a} &(7.40d) \cr }$$
and the field equations are unchanged.  In terms of the natural conformal
expressions ${\cal {R}}_{a}$, ${\cal {T}}_{a}$ and $\omega_{0}^{0}$ the
field equations are
$$\eqalignno{
{\cal {R}}_{a} &= {\cal {T}}_{a} &(7.41a) \cr
*{\bf {d}}*{\bf {d}}\omega_{0}^{0} &= {\bf {J}}  &(7.41b) \cr }$$

The expressions for the curvature above make use of the tracelessness
condition $\Omega_{bac}^{a}= 0$ to write the co-solder form in terms of the
Ricci tensor instead of the stress-energy tensor.  Considering the form of
the curvature if we do not impose the tracelessness condition, we can
interpret the Einstein equation as being that condition that reduces
$\Omega_{b}^{a}$ and $\Omega_{a}^{0}$ to the Weyl curvature and its
divergence, respectively.

It should be pointed out that eqs.(7.40) are quite remarkable for
biconformal curvatures in that none of the curvatures has a cross-term or
momentum-term.  Despite the 8-dimensional formulation of the theory, all
terms containing ${\bf {d}}y_{\mu}$ have dropped out.  The space spanned by
the $y$-coordinates is therefore flat, and plays no role in the
gravitational effects of this class of models.

In the final section, we discuss the observability of the extra dimensions
of biconformal spaces.
\bigskip
\noindent 8.  Discussion
\medskip
By placing conformal gauging on an 8-dimensional base space instead of the
usual 4-dimensional base space, we have overcome the long-standing problem
of size change in physical models based on scale-invariance. In this
section, we demonstrate that our interpretation of biconformal gauge theory
is consistent with experience.  In particular, we examine the observable
consequences of the added dimensions with regard to structure and function.

Three techniques have been used for adding extra dimensions in fundamental
models of the world:  (a) topological compactification of the extra
dimensions, (b) construction of laws of motion or field equations that
dynamically reduce the extra dimensions  to a sufficiently small scale that
they play no macroscopic role [48], or (c) identification of the extra
dimensions with everyday properties already associated with macroscopic
matter.   Technique (a) is routinely employed in Kaluza-Klein field
theories, while technique (b) has been used recently [49, 50] in an attempt
to associate a fifth dimension with mass.  The third technique, (c), was
used in the development of special relativity where time came to be seen as
a coordinate in a higher (i.e., four) dimensional space rather than as a
parameter of an intrinsically different type on a three-dimensional space.
The local interpretation of biconformal space is of type (c).  One should
therefore {\it {not}} think in terms of compactification or other standard
Kaluza-Klein ideas.  Instead, we regard the extra four dimensions of the
biconformal co-space as familiar, routinely observable macroscopic
dimensions, and ask whether a coordinate interpretation of 4-momentum is
consistent with experience.  We demonstrate this consistency by focusing on
the two necessary axes of correspondence:
\smallskip
\item{A.} Intrinsic structure

\leftskip=.5in
\rightskip=.3in
\noindent The biconformal co-space must have the same mathematical
structure and transformation properties as momentum space.

\leftskip=0in
\rightskip=0in

\item{B.} Dynamical function

\leftskip=.5in
\rightskip=.3in
\noindent The biconformal dynamical laws and description of collisions or
interactions must accord with experience.

\leftskip=0in
\rightskip=0in

\noindent We address each of these in turn, then conclude by citing
positive evidence for a coordinate interpretation of momentum.

\medskip
\noindent A. Intrinsic structure
\smallskip
For either flat or curved spacetimes, momentum space is the tangent space
at each point of spacetime.  To identify the biconformal co-space with the
tangent space we must check that both the Minkowski vector space structure
of the tangent space and transformation properties of the tangent basis are
reflected in the co-space.

The biconformal co-space is a Minkowski vector space if and only if it is
flat.  This flatness is obvious for the flat solutions studied in Secs.(3)
- (6), while is a consequence of the vanishing torsion and dilation in the
curved models of Sec.(7).  In either of these classes it is therefore
possible to set up a vector space isomorphism between the tangent space and
the co-space.  We note that the same isomorphism also holds between the
tangent space of a Riemannian spacetime and the co-space coordinates of a
{\it {normal biconformal space}} [51].  Normal spaces are defined to be
torsion-free spaces in which the dilational curvature is closed, the Weyl
1-form is exact, and the spacetime term of the curvature tensor is
trace-free.  We therefore have two large, disjoint classes of biconformal
geometries in which the biconformal co-space is isomorphic to the
Riemannian tangent space.

There are two transformation properties of momentum variables which we must
also check.  First, under Lorentz transformations, $p_{\mu}$ transforms
with the inverse to the {\it {same}} transformation as $x^{\mu}$ in flat
spaces or ${\bf {d}}x^{\mu}$ in curved spaces.  That is, there must not be
two independent Lorentz transformations which can be applied to the space
and co-space separately.  This is a notable property of the 8-dimensional
gauging of the 4-dimensional conformal group.  Because we began with the
spacetime conformal group, there is only a single set of local Lorentz
transformations.  Moreover, $y_{\mu}$ has the proper covariant form.  The
second transformation property required of momentum is that its scaling
weight be $-1$, since (using Planck's constant) momentum has geometric
units of inverse length.  This corresponds correctly to the inverse length
units of $y_{\mu}$.
\medskip
\noindent B. Dynamical function
\smallskip
We now consider whether Hamiltonian dynamical laws and our experience of
collisions or interactions are consistent with a coordinate interpretation
of momentum.  There are three essential points concerning a coordinate
interpretation which must correspond to our usual experience.
\smallskip
\leftskip=.3in
\rightskip=.2in

\noindent \item{1.} Hamiltonian dynamical equations should describe the
classical motion of particles and fields.

\noindent \item{2.} While there is no continuity requirement on momenta,
coordinates should not change discontinuously.

\noindent \item{3.} From our experience we know that collisions only occur
between particles which are nearby in their spacetime coordinates, and do
{\it {not}} necessarily occur when their momentum separation becomes small.
We must show how this observation is consistent with the single proper
biconformal separation, which depends on both position separation and on
momentum separation.

\leftskip=0in
\rightskip=0in
\smallskip

For agreement on point 1, we require a match of dynamical properties
between the Hamiltonian dynamics of momentum space and the dynamics of
biconformal spaces.  In Secs.(3)-(6), we showed that the usual Hamiltonian
dynamical picture is a natural property of flat biconformal spaces.  Not
only do the co-space coordinates act as momenta, but also the symplectic
form given by the exterior derviative of the Weyl 1-form\footnote{$^3$}{The
structure equation ${\bf {d}}\omega_{0}^{0} = \omega_{0}^{a}
\omega_{a}^{0}$ shows that ${\bf {d}}\omega_{0}^{0}$ is a manifestly
closed, nondegenerate 2-form, hence symplectic.} provides the usual
Hamiltonian dynamical structure associated with 1-particle phase space.
Indeed, we demonstrate that any Hamiltonian system generates a
super-Hamiltonian hypersurface in a unique flat biconformal space, and
conversely that a hypersurface in a flat biconformal space gives a unique
Hamiltonian system.

In the dilation-free curved geometries of Sec.(7) and in normal biconformal
spaces we can expect the Hamiltonian dynamical laws of flat biconformal
space to hold {\it {locally}}.  As long as we look at a sufficiently small
neighborhood to permit a single particle picture, the considerations of
Secs.(3) - (6) hold without modification.  That fields on spacetime also
pose no problem may be seen from three different perspectives:

\leftskip=.5in
\rightskip=.3in
\item{(a)} The observable properties of fields are characterized by tensors
built on the tangent space, so the isomorphism between the tangent space
and the co-space insures that the same properties can be measured in both
models.

\item{(b)} Fields and their properties are derivable as the continuum limit
of many individual particles, so as long as interactions between particles
are correctly predicted, the field limit may be expected to hold.

\item{(c)}  Fields must admit classical, single particle limits.  Thus, the
highly localized field of an isolated electron moves according to a
classical single particle Hamiltonian.

\leftskip=0in
\rightskip=0in

The net effect of these considerations is that there is no particle or
field property that one might observe in a flat, dilation-free or normal
biconformal space that could not also be described in the tangent bundle of
a 4-dimensional Riemannian geometry.  Moreover, biconformal spaces
automatically predict the presence and form of electromagnetic fields and
the Lorentz force law, properties which in a Riemannian spacetime must be
added by hand.
\smallskip
Now consider point 2.  The nearly instantaneous change of momentum which
occurs when, say, a ball bounces off a wall seems to violate the notion of
continuous motion in the momentum dimensions.  However, the continuity of
the actual dynamics underlying the bounce can be seen in either of two
ways.  Most simply, we can choose the superhamiltonian with a
phenomenological potential representing the wall.  A realistic potential
will not have an actual discontinuity, so the resulting motion predicted by
Hamilton's equations (or by the biconformal involution of eqs(4.5b, c))
will be continuous.

A more fundamental way of seeing the continuity of momentum during the
collision is to use the electrodynamic law predicted by local biconformal
theory to examine the actual motion of each constituent particle of the
ball in the appropriate background field.  In this view, the apparent
discontinuity arises because the region of biconformal space chosen for
study is too large for the flat approximation to be valid.  Shrinking the
region to one where a single particle interpretation is expected to hold
solves the problem.
\smallskip
Moving to point 3, we know that in order to collide or interact strongly,
the proper separation of two particles must be small.  For this fact to be
consistent with our experience that collisions occur whenever the spacetime
separation becomes small regardless of the momentum separation, is
nontrivial.  Indeed, recent models [49, 50] in which a fifth dimension
proportional to mass is added to spacetime fail this test $-$ in those
models, two particles of vastly different mass will not generally collide.

In the biconformal models presented here, we find that the proper distance
between colliding objects behaves according to our experience.  The proper
inverval is given by conformal Killing metric of eq.(6.8).   While eq.(6.8)
applies to the orthonormal $(\omega_{0}^{a}, \omega_{a}^{0})$ basis, we
easily change to the $(x^{a}, y_{a})$ coordinate basis and find that the
metric takes the form
$$g_{ab} = \pmatrix{-2{\cal {W}}_{(ab)} & \delta_{m}^{\enskip n} \cr
\delta_{n}^{\enskip m}& 0 \cr }$$
where ${\cal {W}}_{ab} = {\cal {T}}_{ab} +  \alpha_{a,b} + W_{a}W_{b} -
{\scriptstyle {1 \over 2}}W^{2} \eta_{ab}$.  Therefore the squared interval
$ds^{2}$ between the ball and the wall in terms of their phase space
separations $(\Delta x^{m}, \Delta p_{n})$ is
$$ds^{2} = -2{\cal {W}}_{mn}\Delta x^{m} \Delta x^{n} + 2 \Delta x^{m}
\Delta p_{m}$$
which vanishes when the spacetime separation $\Delta x^{m}$ vanishes and
does not necessarily vanish when the momentum separation $\Delta p_{m} =
0$.  This is exactly what is required.
\bigskip
In parts A and B we have shown that there is no mystery to the extra four
dimensions of the biconformal co-space.  Indeed, the coordinates for these
extra directions are always immediately available as the energy and
momentum of the system under study.  Thus, if we want to probe the full
biconformal geometry by walking off in the extra directions, we already
know exactly how to do it.  We can simply vary our energy or the direction
of our motion.  Any change of the tangent to our world line is a change in
location in the momentum sector of biconformal space.

The new picture would be particularly convincing if there were some direct
positive evidence that (perhaps under extreme conditions) the full
8-dimensional character of the world is manifest.  For such evidence we
must study biconformal spaces in which the fields take on a more general
$y$-dependence. Therefore, the following observations are necessarily
conjectural, the investigation of the microscopic meaning of biconformal
space lying clearly beyond the scope of the present paper.  Nonetheless, as
long as the deviation from the normal or dilation-free classes are small,
we may expect the new evidence to show up as a dependence of some physical
fields or parameters on momentum as well as position.  We briefly discuss
two physical variables which have this property: quantum mechanical wave
functions, and running coupling constants.

The most evident dependence of fundamental physics on momentum is the phase
space duality found in quantum mechanics.  It is well understood that the
wave function of a particle may be equally well represented using either
momentum or position coordinates.  In a very direct sense, the momentum of
the particle is used as a coordinate.  Representations such as the number
basis for the harmonic oscillator, which lies midway between momentum and
position variables, are also frequently used.

The fact that we do not independently probe both position and momentum
simultaneously suggests that even at the quantum level (at least
semi-classically) matter is substantially restricted to the neighborhood of
some phase-space hypersurface.  This might be understood in terms of
biconformal space as near-normal biconformal behavior, with microscopic
deviations from an involute or nearly involute subspace.  The picture is
consistent with the view of the path integral approach, which says that a
quantum system essentially probes all phase space paths.  In the
biconformal picture, a path integral would simply be taken over all eight
coordinates.

A second piece of evidence that momentum components act as independent
coordinates is the existence of running coupling constants in quantum field
theory.  It is found experimentally that at high energies, the strength of
the electromagnetic and weak couplings vary with energy.  Such
energy-momentum dependence is easily understood if energy and momentum are
coordinates, but is otherwise a somewhat non-transparent result of detailed
field theoretic calculation.

It remains to be seen whether specific biconformal models can be developed
which will make these final observations precise.

\vfil
\break
\hfil References
\smallskip
\item{[1]} Weyl, H., Sitzung. d. Preuss. Akad. d. Wissensch. (1918) 465;
The Principle
of Relativity, Chapter XI, (Dover, 1923).
\item{[2]} Weyl, H., {\it {Space-Time-Matter}}, Dover Publications, New York
(1952).  Originally:  H. Weyl,  Raum-Zeit-Materie, (3rd Ed.,1920), Chapts
II \& IV,
§§34+35, p.242,et seq.
\item{[3]} Weyl, H., Nature 106 781(1921); Math. Zeitschr.  2 (1918) 384;
Ann. d.
Physik,  54  (1918)117 ; Ann. d. Physik,  59 (1919)101; Phys. Zeitschr. 21
(1920)649;
Ann. d. Phys. 65 (1921)541; Phys. Zeitschr. 22 (1921)473; Zeit. f. Physik, 56
(1929)330.
\item{[4]} Pauli, W., {\it {Theory of Relativity}} Translated by G. Field,
Dover
Press, New York (1958) 192
\item{[5]} A. Einstein, S.B. preuss. Akad. Wiss. 478 (1918), including
Weyl's reply.
\item{[6]} F. London,  Zeitschr. f. Physik 42 375 (1927).
\item{[7]} P.A.M.Dirac, Proceedings of the Royal Society, A209, 291 (1951).
\item{[8]} P.A.M.Dirac, Proceedings of the Royal Society, A212, 330 (1952).
\item{[9]} Pauli, W., {\it {Theory of Relativity}} Translated by G. Field,
Dover
Press, New York (1958) 192
\item{[10]} Adler, R., M. Bazin, M. Schiffer, {\it {Introduction to General
Relativity}}, McGraw Hill, New York (1965) 401.
\item{[11]} J. Ehlers, A.E.Pirani, and A.Schild, in {\it {General
Relativity}},  edited
by L. O'Raifeartaigh (Oxford University, Oxford, 1972).
\item{[12]} Utiyama, R., Prog. Theor. Phys. {\bf {50}} (1973) 2080.
\item{[13]} P.A.M. Dirac, Proc. Roy. Soc. London  A333 (1973) 403.
\item{[14]} Freund, P. G. O., Ann. Phys. {\bf {84}} (1974) 440.
\item{[15]} Utiyama, R., Prog. Theor. Phys. {\bf {53}} (1975) 565.
\item{[16]} P.G. Bergmann, {\it {Introduction to the Theory of
Relativity}}, Chap
XVI (Dover, 1976).
\item{[17]} Hayashi, K., M. Kasuya and T. Shirafuji, Prog. Theor. Phys.
{\bf {57}}
(1977) 431.
\item{[18]} Hayashi, K. and T. Kugo, Prog. Theor. Phys. {\bf {61}} (1979) 339.
\item{[19]} J. Audretsch,  Phys. Rev. D27 2872 (1983).
\item{[20]} J. Audretsch, F. Gähler and N. Straumann,  Commun. Math. Phys.
95, 41
(1984).
\item{[21]} Ranganathan, D. , J. Math Phys. {\bf {28}} (1986) 2437.
\item{[22]} Cheng, H. Phys. Rev. Lett. {\bf {61}} (1988) 2182.
\item{[23]} Ferber, A. and P.G.O. Freund, Nucl. Phys. {\bf {B122}} (1977) 170.
\item{[24]} Crispim-Rom{\~a}o, J., A. Ferber and P.G.O. Freund, Nucl. Phys.
{\bf
{B126}} (1977) 429.
\item{[25]} Kaku, M., P.K. Townsend and P. Van Nieuwenhuizen, Phys. Lett. {\bf
{69B}} (1977) 304.
\item{[26]} F. Mansouri, Phys. Rev. Lett. 42 (1979) 1021.
\item{[27]} F. Mansouri and C. Schaer, Phys. Lett. 101B (1981) 51.
\item{[28]} Wheeler, J. T., Phys Rev D{\bf {44}} (1991) 1769.
\item{[29]} Cartan, \'{E}., La th\'{e}orie des groupes finis et continus et la
g\'{e}om\'{e}trie diff\'{e}rentielle, Paris, Gauthier-Villars (1937);
Misner, C.W, K. S.
Thorne and J. A. Wheeler,  {\it {Gravitation}}, W. H. Freeman and Co., San
Francisco (1970) and refs. therein.
\item{[30]} Klein, F., ``Erlangerprogram: Vergleichende Betrachtungen
\"{u}ber neuere
geometrischen Forschungen" (1872), translated in Bull. Amer. Math. Soc. 2
(1893) 215-
249.
\item{[31]} Steenrod, N., {\it {The Topology of Fibre Bundles}}, Princeton
University Press, Princeton, 1951.
\item{[32]} Kobayashi, S. and K. Nomizu, {\it {Foundations of Differential
Geometry}} Wiley, New York, 1963.
\item{[33]} Utiyama, R., Phys. Rev. {\bf {101}}, (1956) 1597.
\item{[34]} Kibble, T.W.B., J. Math. Phys., {\bf {2}}, (1961) 212.
\item{[35]} S.W. MacDowell and F. Mansouri, Phys. Rev. Lett. 38 (1977) 739.
\item{[36]} Freund, P.G.O., {\it {Introduction to Supersymmetry}}, Cambridge
University Press, Cambridge.  See Chapter 21, pp 99-105.
\item{[37]} Chevalley, C., {\it {Theory of Lie Groups}}, Princeton Unviversity
Press (1946).
\item{[38]}  Clifton, Y. H., private communication.
\item{[39]} Eguchi, T., P. B. Gilkey and A. J. Hanson, Phys. Rep. {\bf
{66}}, (1980)
213.
\item{[40]} Hughston, L. P. and T. R. Hurd, Physics Reports 100, No.5
(1983) 273 -
326.
\item{[41]} Wheeler, J. T., to be published in  the proceedings of the
Seventh Marcel
Grossman conference.
\item{[42]} Szekeres, P., Proc. Roy. Soc. London, {\bf {A274}}, (1963) 206
- 212.
\item{[43]} Misner, C. W., K. S. Thorne  and J. A. Wheeler, {\it
{Gravitation}},
W. H. Freeman and Company, San Francisco (1973) 488.
\item{[44]} Santamato, E., Phys. Rev. D, {\bf {29}} (1984) 216.
\item{[45]} Caianiello, E. R., M. Gasperini, E. Predazzi and G. Scarpetta,
Phys. Lett.
{\bf {132A}}, 2, (1988) 82
\item{[46]} Wheeler, J. T., Phys Rev D{\bf {41}} (1990) 431.
\item{[47]}  Caianiello, E. R., A. Feoli, M. Gasperini and G. Scarpetta,
Intl. J. Theor.
Phys., {\bf {29}}, 2, (1990), 131 and references therein.
\item{[48]} Gell-Mann, M., Cal. Tech. field theory seminar (1985).
\item{[49]} Mashhoon, B., H. Liu and P. Wesson, Phys. Lett. B{\bf {331}}
(1994) 305.
\item{[50]} Liu, H. and B. Mashhoon, Ann. Physik {\bf {4}} (1995) 565.
\item{[51]} Wheeler, J. T., submitted for publication.
\bye